\pdfoutput=1
\documentclass[lettersize,journal]{IEEEtran}
\usepackage{amsmath,amsfonts}
\usepackage{mathrsfs}
\usepackage{algorithmic}
\usepackage{array}
\usepackage{subfigure}
\usepackage[caption=false,font=normalsize,labelfont=sf,textfont=sf]{subfig}
\usepackage{color}
\usepackage{textcomp}
\usepackage{stfloats}
\usepackage{url}
\usepackage{verbatim}
\usepackage{graphicx}
\graphicspath{{figures/}}
\usepackage{multirow}
\usepackage{diagbox}
\usepackage{booktabs}
\usepackage{cite}
\def\BibTeX{{\rm B\kern-.05em{\sc i\kern-.025em b}\kern-.08em
		T\kern-.1667em\lower.7ex\hbox{E}\kern-.125emX}}
\usepackage{balance}

\begin{document}
	\title{Spatial-temporal Graph Based Multi-channel Speaker Verification With Ad-hoc Microphone Arrays}
	\author{Yijiang Chen, Chengdong Liang, and Xiao-Lei Zhang
		\thanks{Yijiang Chen and Xiao-Lei Zhang are with the School of Marine Science and Technology, Northwestern Polytechnical University, 127 Youyi West Road, Xi'an, Shaanxi 710072, China (e-mail: orangechen@mail.nwpu.edu.cn, xiaolei.zhang@nwpu.edu.cn). }
		\thanks{Chenegdong Liang is currently with the Horizon Robotics, Beijing, China. The work was done when Chengdong Liang was with the Northwestern Polytechnical University, China (e-mail: chengdong01.liang@horizon.ai). }

}

	\markboth{Journal of \LaTeX\ Class Files,~Vol.~18, No.~9, September~2020}%
	{How to Use the IEEEtran \LaTeX \ Templates}
	
	\maketitle

	\begin{abstract}
		The performance of speaker verification degrades significantly in adverse acoustic environments with strong reverberation and noise. To address this issue, this paper proposes a spatial-temporal graph convolutional network (GCN) method for the multi-channel speaker verification with ad-hoc microphone arrays. It includes a feature aggregation block and a channel selection block, both of which are built on graphs. The feature aggregation block fuses speaker features among different time and channels by a spatial-temporal GCN. The graph-based channel selection block discards the noisy channels that may contribute negatively to the system. The proposed method is flexible in incorporating various kinds of graphs and prior knowledge. We compared the proposed method with six representative methods in both real-world and simulated environments.
		Experimental results show that the proposed method achieves a relative equal error rate (EER) reduction of $\mathbf{15.39\%}$ lower than the strongest referenced method in the simulated datasets, and $\mathbf{17.70\%}$ lower than the latter in the real datasets. Moreover, its performance is robust across different signal-to-noise ratios and reverberation time.
	\end{abstract}
	
	
	\section{Introduction}
	\IEEEPARstart{S}{peaker} verification is to identify whether a speaker is the target speaker. It finds important applications in privacy protection, identity authentication, smart home, etc. The research on speaker verification dates back to the 1960s \cite{Pru1964}, followed by a series of statistical-model-based approaches, such as the Gaussian-mixture-model-based universal background model (GMM-UBM) \cite{REYNOLDS200019} and i-vectors \cite{ivector}. With the rise of deep learning era, speaker feature extraction with neural networks becomes the mainstream \cite{Snyder2017,xie2019,muck2018}. Although the deep-learning-based speaker verification has achieved a significant breakthrough, far-field speaker verification is still challenging. When a microphone is placed far away from a speaker, the recorded speech signal is not only severely attenuated but also corrupted by background noise, reverberation and other interfering sound sources. Eventually, the performance of a speaker verification system in the far-field conditions drops sharply.
	
	To compensate the negative effect caused by noise and reverberation, a common approach for speaker verification is to add a deep-learning-based speech noise reduction front-end \cite{zhao2019robust,shon2019voiceid,kolboek2016speech,chang2017robust,michelsanti2017conditional,plchot2016audio}. For example, Kolboek \emph{et al.} \cite{kolboek2016speech} first used a masking-based front-end to compute the posterior probability of a frame vector belonging to a speaker for the GMM-UBM-based speaker verification system. Then, Chang and Wang \cite{chang2017robust} used long short-term memory as a masking-based front-end to estimate clean speech for speaker recognition. Novotny \emph{et al.} \cite{novotny2018use} learned a mapping from noisy speech to clean speech using an encoder, and subsequently applied the estimated speech to extract segment-level speaker representations. Another class of approaches treat far-field speaker recognition as a domain mismatch problem. It regards clean speech as the source domain and noisy data as the target domain, and solves the domain mismatching problem by domain adaptation methods \cite{meng2019,Peri2020,Qin2019}. The aforementioned methods are based on a single-channel microphone, which does not explore important spatial information.
	
	To address this issue, multi-channel speaker verification based on fixed arrays utilize azimuth information for further performance improvement. Taherian \emph{et al.} \cite{taherian2019deep} used a deep-learning-based minimum variance distortionless response to get the enhanced speech for speaker verification, where the deep neural network is used to learn a time-frequency mask for estimating the noise component of each channel. Then, they further explored the effect of combining different beamforming front-ends with i-vector/x-vector-based recognition back-ends. Yang and Chang \cite{yang2019joint} jointly optimized the deep-learning-based beamforming and speaker verification. Cai \emph{et al.}\cite{cai2019multi} took the multichannel noisy speech as the input of a two-dimensional-convolutional-neural-network-based end-to-end speaker recognition directly, which yields lower equal error rate (EER) than single-channel speaker recognition systems. He \emph{et al.} \cite{he2021multi} extracted vocal pattern information and orientation information simultaneously by a multi-channel front-end, and applied the orientation information to a direction-of-arrival (DOA) estimation for speaker identification. Similar works which combine other orientation information for speaker identification were also developed in \cite{wang2022spatial,kataria2020multi}. Wang \emph{et al.}\cite{kataria2020multi} obtained spatially encoded s-vectors by DOA estimation of the multichannel input, and identified speakers according to the similarity matrix of the s-vectors and x-vectors. It is worthy noting that the above multi-channel methods used fixed arrays with small array apertures. When a speaker is far from the array, then serious signal attenuation and strong interference of reverberation are still hard to prevent.
	
	To reduce the occurrence probability of extremely hard far-field problems, grouping multiple distributed devices together as an \textit{ad-hoc microphone array} becomes a new type of effective methods, where each device in the ad-hoc microphone array, denoted as an \textit{ad-hoc node}, contains either a single microphone or a conventional fixed microphone array. Compared to the fixed arrays, a key advantage of the deep-learning-based ad-hoc microphone array processing is that it is able to utilize spatial distance information via channel reweigting and selection. An early work used a deep neural network to estimate the signal-to-noise ratio at each randomly placed microphone array for the channel reweighting and selection \cite{zhang2021deep}. However, the channel selection approach is not optimized. Recently, many works explored advanced channel selection approaches for speech enhancement \cite{TADRN,fasnet}, speech separation \cite{yang2022deep,wang2021continuous,tac}, speech recognition \cite{chen2021scaling}, and speaker recognition \cite{liang2021attention,cai2021embedding}. Particularly, Liang \textit{et al.} \cite{liang2021attention} and Cai \textit{et al.} \cite{cai2021embedding} independently proposed end-to-end speaker verification with ad-hoc microphone arrays, where an inter-channel attention-based channel reweighting method was developed to fuse utterance-level speaker features from all channels. However, the spatial-temporal connection between the nodes were not fully explored by simply the attention mechanism. Furthermore, because the ad-hoc nodes that are far away from speech sources might be too noisy to contribute negatively to the system, taking all channels into account may not be the best choice. Recently, some methods explored graphs to reweight and fuse the multi-channel signals collected by distributed microphone arrays for speech enhancement \cite{jacob2021icassp,Haostgraph}. However, they simply used complete graphs without further exploring different kinds of graphs that can incorporate flexible prior knowledge. Their effectiveness was not studied in speaker verification as well.

	To fully exploit the spatial-temporal information, in this paper, we propose an end-to-end multichannel speaker verification framework based on graph convolutional networks (GCN).
	{It includes a graph-based spatial-temporal multi-channel feature aggregation block and a graph-based channel selection block. The former learns to enhance the speaker characteristics of each frame of a channel by aggregating spatial-temporal information from its neighboring frames and channels, while the latter discards strongly-noisy channels to further improve the performance. The core contributions of the paper are as follows.
		\begin{itemize}
			\item \textbf{A graph-based spatial-temporal aggregation framework is proposed for multi-channel speaker verification with ad-hoc microphone arrays.} Unlike existing works \cite{jacob2021icassp,Haostgraph} which regard microphones as vertices of a graph for noise reduction, the proposed method takes both channels and time frames as vertices of a graph for speaker verification. It can capture the relationship between time frames across channels and thus has stronger modeling capability than simply modeling the relationship between channels.
			\item \textbf{Several spatial-temporal GCNs that not only accelerate the modeling process significantly but also are flexible in incorporating prior knowledge are proposed under the framework.} Unlike the methods \cite{jacob2021icassp,Haostgraph,2021arXiv211005975L} which focus on applying graph neural networks (GNNs) without exploring adjacent matrices of graphs, this paper constructs adjacent matrices of graphs that are flexible in utilizing more abundant prior information, such as the spatial labeling information of datasets, temporal labeling information of speakers. Moreover, this paper describes two efficient backbone spatial-temporal GCNs.
		\end{itemize}
	 Experimental results showed that proposed methods outperform six representative comparison methods significantly in highly-reverberant and low signal-to-noise ratio environments on both a simulated dataset and two real-world datasets.
	This paper differs from our preliminary work \cite{2021arXiv211005975L} in several major aspects, which includes the design of the channel selection block and various methods to construct graphs (but not in \cite{2021arXiv211005975L}), which improves the performance over \cite{2021arXiv211005975L} vitally. Consequently, many new experimental scenarios were studied beyond that in \cite{2021arXiv211005975L}.
	}

	The rest of the paper is organized as follows. Section~\ref{sec:relatedwork} briefly reviews the research progress of graph neural networks and introduces some preliminaries of modeling ad-hoc microphone arrays with graph neural networks.  Section~\ref{sec:framework} introduces the proposed framework. Detailed description of the proposed algorithm is presented in Section~\ref{sec:channel_select}, \ref{sec:aggre_inplementation}, \ref{sec:channel_select} respectively. Section~\ref{sec:exp_setup} describes the experimental settings. Section~\ref{sec:result} reports experimental results. Finally, Section~\ref{sec:conclusion} concludes the paper.
	
	\section{Related work} \label{sec:relatedwork}
	{
	Graph is expressed as $G=(\mathcal{V}, \mathcal{E})$, where $\mathcal{V}$ represents a set of nodes, and $\mathcal{E} \subseteq \mathcal{V} \times \mathcal{V}$ represents a set of edges between nodes. Given a node $v \in \mathcal{V}$, its neighboring nodes are defined as $\mathcal{N}(v) = \{u \in \mathcal{V} | e_{vu} \in \mathcal{E}\}$, where $e_{vu}$ means that the node $v$ is connected with its neighbor node $u$. The graphs with attributes are called attributed graphs. In GNNs, the input of an attributed-graph is expressed as $(\mathbf{X}_{\mathcal{V}}, \mathbf{X}_{\mathcal{E}}, \mathbf{A})$, where $\mathbf{X}_{\mathcal{V}} \in \mathbb{R}^{D_{\mathcal{V}}}$ denotes the attributes of nodes, $\mathbf{X}_{\mathcal{E}}\in \mathbb{R}^{D_{\mathcal{E}}}$ denotes the attributes of edges, $D_{\mathcal{V}}$ and $D_{\mathcal{E}}$ represent the feature dimensions of the nodes and edges respectively. The adjacent matrix $\mathbf{A} \in \mathbb{R}^{\mathcal{|V|} \times \mathcal{|V|}}$ contains the edges between any pairs of nodes in the graph. {If the node $v$ and node $u$ are connected, the $u$th row and $v$th column of $\mathbf{A}$ are set to 1.}}
	
{Existing studies on GNNs can be divided into the following four categories \cite{wu2020comprehensive}: recurrent graph neural networks (RecGNNs), convolutional graph neural networks (ConvGNNs), graph autoencoders (GAEs), and spatial-temporal graph neural networks (STGNNs), all of which can be trained at the node level, edge level or graph level, given $(\mathbf{X}_{\mathcal{V}}, \mathbf{X}_{\mathcal{E}}, \mathbf{A})$ as the inputs}. RecGNNs {learn node representations of graph data iteratively by using the same graph recurrent layer} until the representation reaches a stable resolution. {ConvGNNs iteratively learn node representations using different graph convolutional layers}. It can be further categorized into the spectral-based and spatial-based ones. {The convolution in spectral-based ConvGNNs has strong theoretical basis in graph signal processing \cite{kipf2016semi}, while the graph convolution in the spatial-based methods is defined by the information propagation between the center nodes and its neighbors.} The spatial-based ConvGNNs inherently relate to the spectral-based ones \cite{defferrard2016convolutional,li2015gated}, and are widely used in real-world scenarios in recent years \cite{levie2018cayleynets,hamilton2017inductive,velivckovic2017graph,zhang2018end}. {GAEs encode graph data into latent vectors, and then reconstruct the graphs from the latent representations. Different from the above approaches that are built on \emph{static graphs}, STGNNs
aim to learn the spatial-temporal dependencies of time-varying data over \emph{dynamic graphs}}, which finds applications in action detection \cite{yan2018spatial}, 3D point clouds processing \cite{shi2020point,shu20193d}, and traffic forecasting \cite{HuDSTGNN,guo2019attention,wu2019Graphwavenet}.

{Common types of graphs include:}
	\begin{itemize}
		\item \emph{Directed graphs versus undirected graphs}: According to whether the relationship between a pair of nodes is bi-directional, graphs can be divided into {directed graphs} and {undirected graphs}. {For a \emph{directed graph}, the edge from node $v$ to node $u$ is different from the edge from node $u$ to node $v$, and therefore $\mathbf{A}[v,u]$ is unnecessarily equal to $\mathbf{A}[u,v]$ in the adjacent matrix. A graph is \emph{undirected} if and only if $\mathbf{A}$ is symmetric.}
		
		\item \emph{Dense graphs versus sparse graphs}: {According to the denseness of the edges in a graph, graphs can be classified into {sparse graphs} and \emph{dense graphs}. In a \emph{dense graph}, each node tends to be linked to any other nodes. A special case of dense graphs is the \emph{complete graph}, in which every node takes all other nodes as neighbors. In a \emph{sparse graph}, most nodes are not mutually connected.}

		\item \emph{Static graphs versus dynamic graphs}: According to whether the nodes or edges of a graph change over time, graphs can be classified into static graphs and dynamic graphs. If all variables of a graph do not change over time, then it is a \emph{static graph}. {A graph is \emph{dynamic} if its nodes or edges change over time, which can be denoted as $ G=(\mathcal{V}^{(t)}, \mathcal{E}^{(t)}),\quad \forall t=1,2,\dots, T$, where $\mathcal{V}^{(t)}$ and $\mathcal{E}^{(t)}$ are the nodes and edges respectively of the graph at time $t$. The attributes of the graph can be represented as $(\mathbf{X}^{(t)}_{\mathcal{V}}, \mathbf{X}^{(t)}_{\mathcal{E}}, \mathbf{A}^{(t)})$}.
	\end{itemize}

The proposed method is flexible in incorporating various kinds of graphs. In this paper, we will discuss the applications of the undirected graph, dense graph, sparse graph, and static graph to ad-hoc microphone arrays. Particularly, unlike {STGNNs} \cite{yan2018spatial,shi2020point,shu20193d,HuDSTGNN,guo2019attention,wu2019Graphwavenet} which model time-varying data via dynamic graphs, the proposed method models the time-dependency between frames via static graphs, so that the overall spatial-temporal data can be modeled via static graphs as well. This novel graph modeling method on time-varying data is simpler and has less variables to be estimated than STGNN.

\begin{figure}[t]
	\centering
	\includegraphics[scale=0.4]{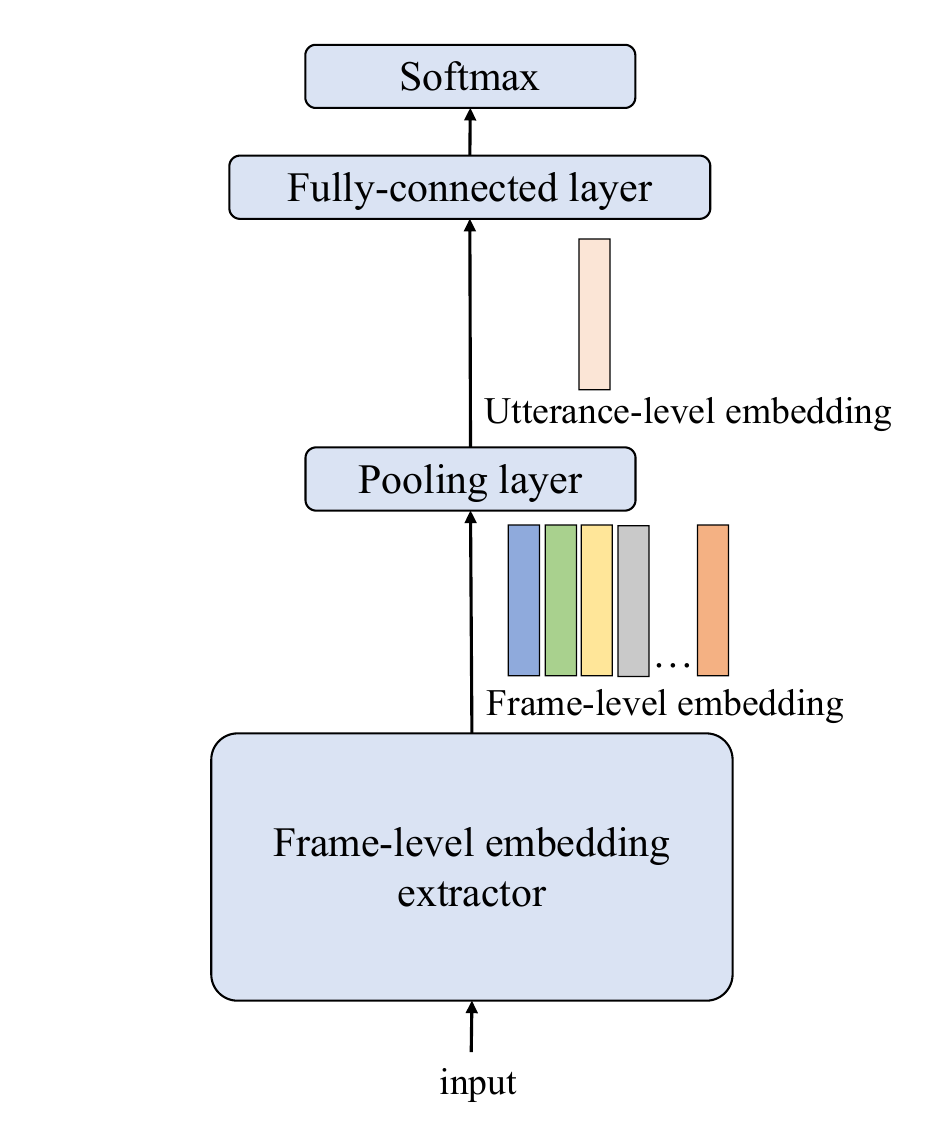}
	\caption{Architecture of the proposed speaker verification system in the first training stage.}
	\label{fig:algorithm_public_single}
\end{figure}

	\section{Framework} \label{sec:framework}

	As shown in Fig.~\ref{fig:algorithm_public_multi}, an ad-hoc microphone array of $C$ randomly distributed devices are placed around a speaker. In this paper, we consider a situation where each device contains a single microphone. With the interference of reverberation and additive noise, the signal collected from the single channel of a device can be formulated as:
	\begin{equation}
		x_{c}(t)=r_{c}(t)*s_{c}(t)+n_{c}(t) ,\quad \forall c=1,2,\ldots,C
	\end{equation}
	where $s_{c}(t)$ and $n_{c}(t)$ denote the clean speech and additive noise of the $c$th channel respectively, the symbol ``$*$'' denotes the convolution operation, and $r_{c}(t)$ denotes the room impulse function (RIR).

\subsection{Two-stage training}

\begin{figure}[t]
	\centering
	\includegraphics[scale=0.4]{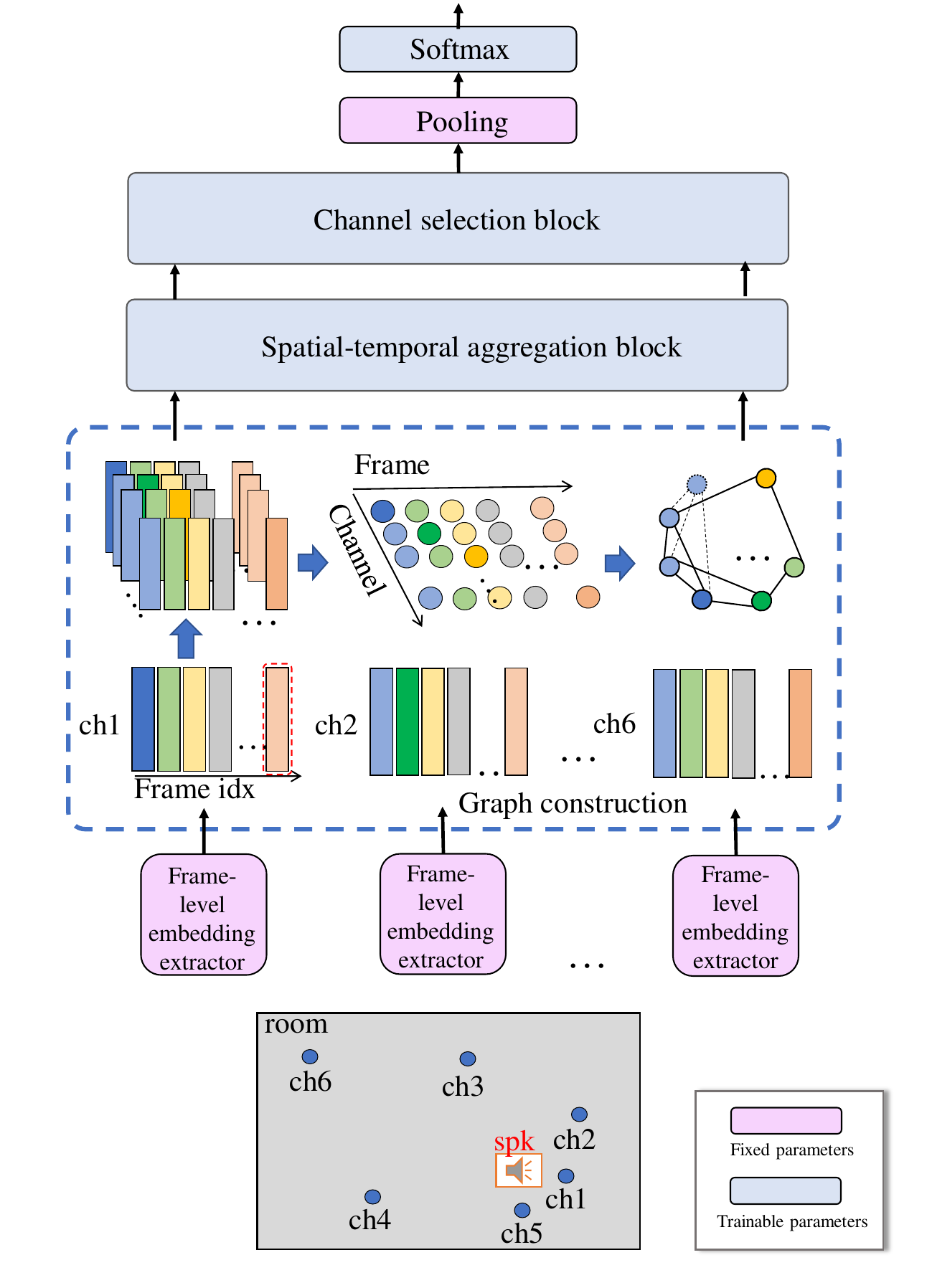}
	\caption{Architecture of the proposed speaker verification system in the second training stage.}
	\label{fig:algorithm_public_multi}
\end{figure}

Given that the data collected from single-channel devices are much more sufficient than that from ad-hoc microphone arrays, we adopt a two-stage training strategy as in \cite{wang2019stream}, so as to prevent the model overfitting to the small-scale data collected from the ad-hoc arrays. In both stages of training, {softmax} is used as the output layer, and  Mel-filterbanks are used as the acoustic features.

The \emph{first-stage training} aims to train a frame-level feature extractor. Specifically, we first train a standard speaker verification system using a large number of single-channel speech data, as shown in Fig.~\ref{fig:algorithm_public_single}. Then, we retain the frame-level feature extractor, and discard all other part of the system.

The \emph{second-stage training} is to train the channel fusion module in Fig.~\ref{fig:algorithm_public_multi} using the spatial-temporal speech data collected from ad-hoc microphone arrays, with the frame-level feature extractor trained in the first stage fixed. Specifically, the frame-level feature extractor is first applied to each channel of the ad-hoc microphone arrays, which generates the frame-level speaker embeddings of each channel:
	\begin{equation}
		\mathbf{X}_{c} \in \mathbb{R}^{T \times D}, \quad c=1,2,\ldots,C
	\end{equation}
	where $D$ is the dimension of the frame-level speaker embeddings, and $T$ is the number of frames. {The speaker embeddings of all $C$ channels can be represented as:}
	\begin{equation}\label{eq:xct}
		{\mathbf{X}_{\mathrm{CT}}} = \{\mathbf{X}_{1},\ldots,\mathbf{X}_{C}\} \in \mathbb{R}^{C\times T \times D}
	\end{equation}

\subsection{Static graph formulation of spatial-temporal data}
In the second-stage training, there are many ways to formulate $\mathbf{X}_{\mathrm{CT}}$ as a graph data, however, this problem seems far from explored yet. To our knowledge, existing works \cite{jacob2021icassp,Haostgraph} do not fully explore the temporal connections between frames. To address this issue, one possible way is to formulate ${\mathbf{X}_{\mathrm{CT}}}$ as a dynamic graph defined in Section \ref{sec:relatedwork} directly. However, this way is too complicated since that both the nodes, edges, and adjacent matrix of a dynamic graph are time-varying.

To prevent this overcomplicated formulation, we propose to reformulate each frame of the spatial-temporal data $\mathbf{X}_{\mathrm{CT}}$ as a node of a static graph, and define its adjacent matrix $\mathbf{A}_{\mathrm{CT}}$ as a boolean matrix:
	\begin{equation}\label{eq:a2}
		\mathbf{A}_{\mathrm{CT}} \in \mathbb{B}^{(C T)\times (C T)}
	\end{equation}
	where its element $A_{\mathrm{CT}}[i,j] = 0$ means that the $i$th node does not have a direct connection with the $j$th node. To this end, we have formulated $\mathbf{X}_{\mathrm{CT}}$ as a static graph $G_{\mathrm{CT}} = \{\mathbf{X}_{\mathrm{CT}}, \mathbf{A}_{\mathrm{CT}}\}$.

{To our knowledge, this is the first time that the spatial-temporal data is formulated as a static graph learning problem. This formulation not only can still grasp the spatial-temporal dependency, such as the relative time delay and SNR differences between the channels, but also is easily trained with common graph neural networks.}

One difficulty of the above formulation is that $\mathbf{A}_{\mathrm{CT}}$ is very large, which causes high computational and storage complexities. For example, a speech signal of 10 seconds collected with 40 ad-hoc nodes has an $\mathbf{A}_{\mathrm{CT}}$ of as large as $40000\times 40000$, if the frame-shift is 10 milliseconds. How to approximate $\mathbf{A}_{\mathrm{CT}}$ efficiently is one of the core issues, which will be introduced in Section \ref{sec:aggregation_framework}.

\subsection{Network architecture}
To grasp the spatial-temporal dependency between the frames in $\mathbf{X}_{\mathrm{CT}}$, we design a \textit{graph-based spatial-temporal aggregation block} $\mathscr{H}^{1}(\cdot)$ to transform $\mathbf{X}_{\mathrm{CT}}$ to another multichannel feature $\mathbf{Z}_{\mathrm{CT}} \in \mathbb{R}^{(CT)\times D}$:
	\begin{equation}\label{eq:channel_aggregation}
		\mathbf{Z}_{\mathrm{CT}} = \mathscr{H}^{1}(\mathbf{X}_{\mathrm{CT}},\mathbf{A}_{\mathrm{CT}})
	\end{equation}
See Section \ref{sec:channel_select} for two implementations of $\mathscr{H}^{1}(\cdot)$.

To filter out the channels that contribute negatively to the speaker verification system, we further design a \textit{graph-based channel selection} algorithm $\mathscr{H}^{2}(\cdot)$ to automatically select $K$ channels ($K\leq C$) from $\mathbf{Z}_{\mathrm{CT}}$:
	\begin{equation}\label{eq:channel_select}
		\left(\hat{\mathbf{Z}}_{\mathrm{CT}},\hat{\mathbf{A}}_{\mathrm{CT}}\right) = \mathscr{H}^{2}(\mathbf{Z}_{\mathrm{CT}},\mathbf{A}_{\mathrm{CT}})
	\end{equation}
with $\mathbf{Z}_{\mathrm{CT}}\in \mathbb{R}^{(KT)\times D}$. An important novelty and advantage of the graph-based channel selection is that prior information can be easily injected into $\mathbf{A}_{\mathrm{CT}}$ for the performance improvement. See Section \ref{sec:aggre_inplementation} for the details of $\mathscr{H}^{2}(\cdot)$ in the presence or absence of prior information.

Finally, we calculate the utterance-level speaker embedding $\mathbf{S}$ by the average pooling over all channels and frames of $\hat{\mathbf{Z}}_{\mathrm{CT}}$:
	\begin{equation} \label{eq:finnal}
		\mathbf{S}= \frac{1}{KT} \sum_{i=1}^{KT}{\hat{\mathbf{Z}}_{\mathrm{CT}}[i,:]}
	\end{equation}
which is used as the input of the utterance-level feature extractor.

\section{Graph-based spatial-temporal aggregation block} \label{sec:aggregation_framework}
	
	\begin{figure*}[t]
		\centering
		\includegraphics[scale=0.55]{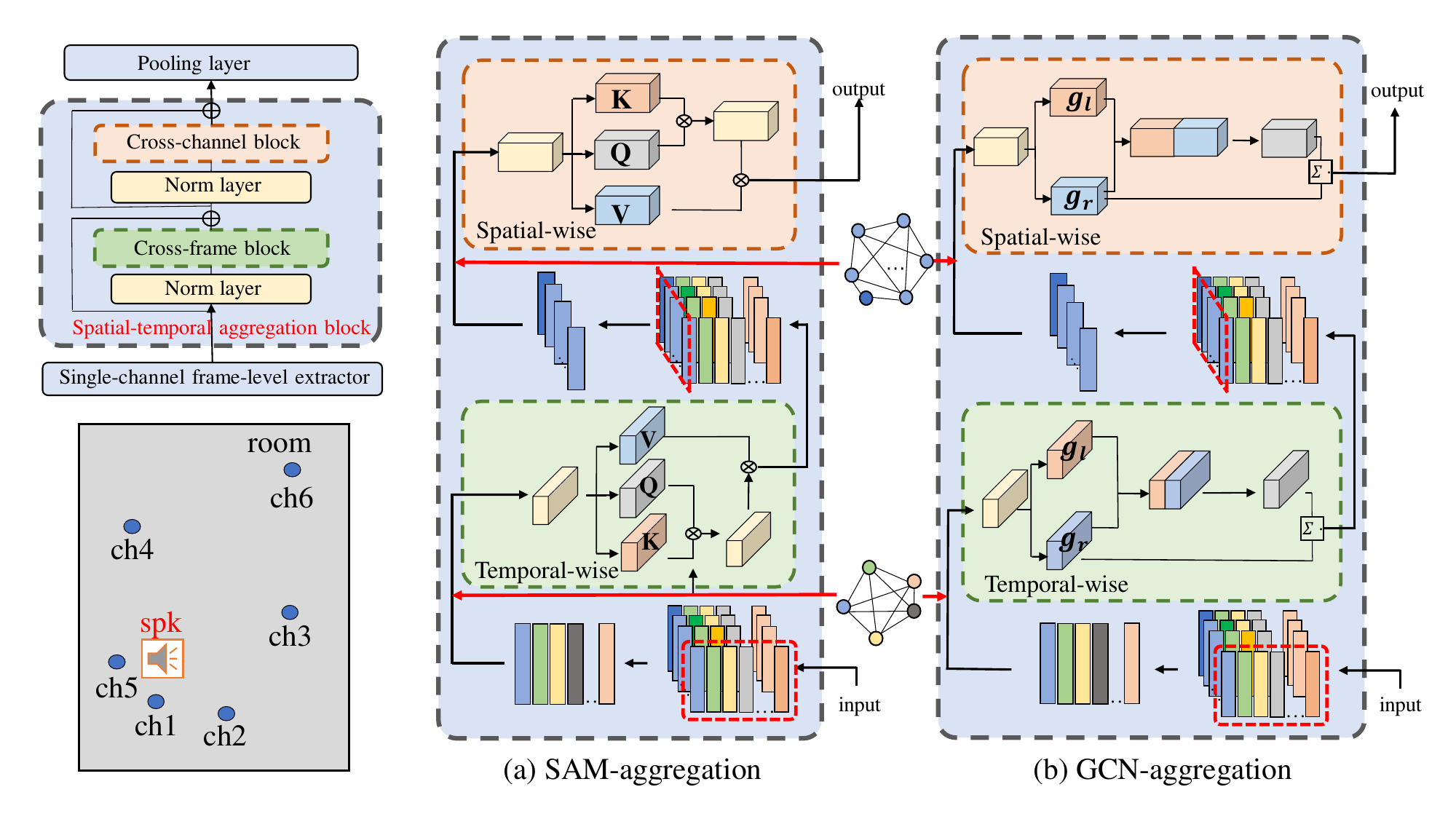}
		\caption{Architectures of two implementations of the graph-based spatial-temporal aggregation block with adjacent matrices.}
		\label{fig:algorithm_public}
	\end{figure*}
	
	To reduce the high computational and storage complexities of the adjacent matrix $\mathbf{A}_{\mathrm{CT}}$ in \eqref{eq:a2}, in this paper, as shown in Fig.~\ref{fig:algorithm_public}, we decompose $\mathscr{H}^{1}(\cdot)$ in \eqref{eq:channel_aggregation} into two successive blocks---a temporal module, denoted as $\mathscr{H}^{1}_{t}(\cdot)$, and a spatial module, denoted as $\mathscr{H}^{1}_{s}(\cdot)$. The overall procedure is as follows:

\subsection{Temporal module}\label{subsec:temporal}

 The temporal module first takes each channel of $\mathbf{X}_{\mathrm{CT}}$ as its input:
	\begin{equation}\label{eq:xct_xc}
		\mathbf{X}^{c}=[\mathbf{x}_{1}^{c},\dots,\mathbf{x}_{t}^{c},\dots,\mathbf{x}_{T}^{c}],\quad \forall c=1,\ldots,C
	\end{equation}
 where $\mathbf{x}_{t}^{c} \in \mathbb{R}^{D}$ is the speaker embedding of the $t$th frame at the $c$th channel. Then, it builds a static graph on each channel, denoted as $G_{\mathrm{temporal}}^c = \{\mathbf{X}^{c}, \mathbf{A}_{\mathrm{temporal}}\}$, where $\mathbf{A}_{\mathrm{temporal}}$ is the adjacent matrix of the temporal static graph, and the frame $\mathbf{x}_{t}^{c}$ is a node of the static graph. Finally, the temporal module is defined as:
\begin{equation} \label{eq:temp_paradigm}
	\mathbf{Y}^{c}= \mathscr{H}^{1}_{t}(\mathbf{X}^{c}, \mathbf{A}_{\mathrm{temporal}}),\quad \forall c=1,\ldots,C
\end{equation}
We then aggregate $\mathbf{Y}^{c}$ into:
\begin{equation} \label{eq:Yct}
	\mathbf{Y}_{\mathrm{CT}}=\{\mathbf{Y}^{1},\dots,\mathbf{Y}^{c},\dots,\mathbf{Y}^{C}\}
\end{equation}
which is used as the input of the spatial module.

A core problem here is the design of $\mathbf{A}_{\mathrm{temporal}}\in \mathbb{B}^{T \times T}$ which is used to constrain the mutual connections between neighboring frames. Because speaker verification requires as many frame-level speaker embeddings as possible for a reliable utterance-level speaker embedding, in this paper, we set $\mathbf{A}_{\mathrm{temporal}}$ to a complete-graph adjacent matrix:
	\begin{equation}
		{A}_{\mathrm{temporal}}[i,j] = 1,\quad \forall i=1,\ldots,T, \mbox{ } \forall j = 1,\ldots, T
	\end{equation}
	
	We also study a common definition in the ablation study of this paper:
	\begin{eqnarray}
		{A}_{\mathrm{temporal}}[i,j] = \left\{\begin{array}{ll}
			1,\quad \mathrm{if} \mbox{ } i\in \mathrm{span}(j,\delta)\\
			0,\quad \mathrm{otherwise}
		\end{array}\right.,
	\end{eqnarray}
	where $\mathrm{span}(j,\delta)= \{j-\delta,\ldots,j-1,j, j+1,\ldots,j+\delta\}$ is a time span of the $j$th frame with a half-window length of $\delta$.

\subsection{Spatial module}

The spatial module first partitions  $\mathbf{Y}_{\mathrm{CT}}$ along the temporal dimension:
\begin{equation} \label{eq:Yt}
	\mathbf{Y}^{t} =[\mathbf{y}^{t}_{1},\dots,\mathbf{y}^{t}_{c},\dots,\mathbf{y}^{t}_{C}],\quad \forall t=1,\ldots, T
\end{equation}
 where $\mathbf{y}_{c}^{t} \in \mathbb{R}^{D}$ represents the frame-level speaker embedding of the $c$th channel at the $t$th frame. Then, it builds a static graph on each frame, denoted as $G_{\mathrm{spatial}}^t = \{\mathbf{Y}^{t}, \mathbf{A}_{\mathrm{spatial}}\}$, where $\mathbf{A}_{\mathrm{spatial}}$ is the adjacent matrix of the spatial static graph, and the frame $\mathbf{y}_{c}^{t}$ is a node of the static graph. Finally, the spatial module is defined as:
\begin{equation} \label{eq:spatial_paradigm}
	\mathbf{Z}^{t}= \mathscr{H}^{1}_{s}(\mathbf{Y}^{t}, \mathbf{A}_{\mathrm{spatial}})
\end{equation}
We aggregate $\mathbf{Z}^{t}$ into:
\begin{equation} \label{eq:finnala}
	\mathbf{Z}_{\mathrm{CT}}= \{\mathbf{Z}^{1},\dots,\mathbf{Z}^{t},\dots,\mathbf{Z}^{T}\}
\end{equation}
 which is used as the input of the graph-based channel selection algorithm.

The adjacent matrix $\mathbf{A}_{\mathrm{spatial}} \ \in \mathbb{B}^{C \times C}$ is used to constrain the mutual connections between the channels of the ad-hoc microphone arrays. Expect for using a complete graph where $\mathbf{A}_{\mathrm{spatial}}$ is an all-one adjacent matrix, we can also implement $\mathbf{A}_{\mathrm{spatial}}$ according to the relative positions between the channels and the speaker:
	\begin{eqnarray} \label{eq:a_spatial}
		{A}_{\mathrm{spatial}}[u,v] = \left\{\begin{array}{ll}
			1,\quad \mathrm{if}  \mbox{ } v \in \mathcal{N}_k(u)\\
			0,\quad \mathrm{otherwise}
		\end{array}\right.,
	\end{eqnarray}
	where $\mathcal{N}_k(u)$ represents the $k$ nearest neighboring nodes of channel $u$. Various methods can be applied to get the $k$-nearest neighborhoods, see Section~\ref{subsec:prior} for the details.

\section{{Implementation of the spatial-temporal aggregation block with adjacent matrices}} \label{sec:aggre_inplementation}

We implemented $\mathscr{H}^{1}_t(\cdot)$ and $\mathscr{H}^{1}_s(\cdot)$ using the same arithmetic mechanism. The difference between $\mathscr{H}^{1}_t(\cdot)$ and $\mathscr{H}^{1}_s(\cdot)$ lies in their inputs and outputs. For clarity, we unify the descriptions of the two functions as follows:
	\begin{equation}
	\mathbf{H} = \mathscr{H}^{1}_{\varepsilon}(\mathbf{X},\mathbf{A}),\quad\forall \varepsilon\in\{t,s\}
	\end{equation}
where $\mathbf{X}\in \mathbb{R}^{N \times D}$ and $\mathbf{A}\in \mathbb{R}^{N \times N}$ denote the node attributes and adjacent matrix of a graph $G$ respectively.

This section describes two implementations of $\mathscr{H}^{1}_{\varepsilon}(\cdot)$, which are the self-attention aggregation with graphs as masks (SAM-agg) and GCN-based aggregation (GCN-agg) respectively. Their structures are shown in the right side of Fig.~\ref{fig:algorithm_public}.

\subsection{Self-attention aggregation with graphs as masks (SAM-agg)}

As showed in Fig.~\ref{fig:algorithm_public}a, SAM-agg aims to learn new attributes $\mathbf{H}$ of the nodes of the graph $G$ by the multi-head self-attention mechanism. When calculating the weight of a node, the adjacent matrix is used to mask off the non-neighbor nodes.

Specifically, suppose we have an $M$-head self-attention model with $M>1$. For the $m$th attention head, we first transform the input $\mathbf{X}$ to the $m$th query $\mathbf{Q}^{m}$, $m$th key $\mathbf{K}^{m}$ and $m$th value $\mathbf{V}^{m}$:
\begin{equation}
	\mathbf{Q}^{m} = \mathbf{X} \mathbf{W}_{Q}^{m}, \mathbf{K}^{m} = \mathbf{X} \mathbf{W}_{K}^{m}, \mathbf{V}^{m} = \mathbf{X} \mathbf{W}_{V}^{m}
\end{equation}
where the matrices $\mathbf{W}_{Q}^{m}$, $\mathbf{W}_{K}^{m}$, and $\mathbf{W}_{V}^{m}$ are learnable parameters of the $m$th attention head, and
 the matrices $\mathbf{Q}^{m}$, $\mathbf{K}^{m}$, and $\mathbf{V}^{m}$ are all in the shape of $\mathbb{R}^{N \times d}$ with $d = E/M$, where $E$ denotes the dimension of the query, key, or value space, which is a user-defined hyperparameter.

The scores between the query-key pair $\mathbf{E}^{m}$ then can be calculated by:
\begin{equation}
	\mathbf{E}^{m}=\frac{\mathbf{Q}^{m}  (\mathbf{K}^{m})'} {\sqrt{d}}
\end{equation}
where $(\cdot)'$ denotes the transpose operator of a matrix.

Given the adjacent matrix $\mathbf{A}$, a local attentive score $\mathbf{\hat{E}}^{m}[n,i]$ from the neighbor node $v_i$ to the central node $v_n$ is calculated by:
\begin{equation}
	\mathbf{\hat{E}}^{m}[n,i] = \frac{\exp{({\mathbf{E}}^{m}[n,i])}}{\sum_{\mathbf{A}[n,j]=1} \exp{({\mathbf{E}}^{m}[n,j])}}
\end{equation}
where the denumerator represents the information from all neighbor nodes of $v_n$ defined by $\mathbf{A}$, i.e. $\{j|v_j \in \mathcal{N}(n),\forall j = 1,\ldots,N\}$.

The hidden {state} of the $m$th head is calculated by:
\begin{equation}
	\mathbf{H}^{m} = \mathbf{\hat{E}}^{m} \mathbf{V}^{m}
\end{equation}

The output of $\mathscr{H}^{1}_{\varepsilon}(\cdot)$ is a concatenation of the hidden states of all attention heads:
\begin{equation}
	\mathbf{H} = \mathrm{concat} [\mathbf{H}^{1},\dots,\mathbf{H}^{m}, \ldots ,\mathbf{H}^{M}]
\end{equation}

\subsection{Graph convolution network based aggregation (GCN-agg)}

As shown in Fig.~\ref{fig:algorithm_public}b, GCN-agg aims to learn new attributes $\mathbf{H}$ by the multi-head self-attention mechanism based on a GCN layer \cite{brody2021attentive}. Different from SAM-agg, the attention mechanism updates the attribute of each node by attending its neighbors using its own representation as the query.

Specifically, the $m$th attention head first projects $\mathbf{X}$ in to a $d$-dimension space using learnable parameters $\mathbf{W}_{l}^{m} \in \mathbb{R}^{D \times d}$ and $\mathbf{W}_{r}^{m} \in \mathbb{R}^{D \times d}$:
\begin{equation}
	\mathbf{g}_{l}^{m} = \mathbf{X}^{c}\mathbf{W}_{l}^{m} ,  \mathbf{g}_{r}^{m} = \mathbf{X}^{c}\mathbf{W}_{r}^{m}
\end{equation}
where $\mathbf{g}_{l}^{m}$ and $\mathbf{g}_{r}^{m}$ denote query and key matrices respectively. The score for the query-key pair is calculated by:
\begin{equation}
	\mathbf{E}^{m}[i,j] = \beta^{\top} \mathrm{LeakyReLU}(\mathrm{concat}(\mathbf{g}_{li}^{m},\mathbf{g}_{rj}^{m}))
\end{equation}
where $\mathbf{E}^{m} \in \mathbb{R}^{N \times N}$, and $\beta \in \mathbb{R}^{2d}$ {is a learnable vector}.

Given the adjacent matrix $\mathbf{A}$, the local attention weight from a neighbor node $v_i \in \mathcal{N}(n)$ to a central node $v_n$, denoted as $\alpha_{ni}^{m}$, is calculated by:
\begin{equation}
	\alpha_{ni}^{m} = \frac{\exp{(\mathbf{E}^{m}[n,i] )}}{\sum_{\mathcal{A}[n,j]=1} \exp{(\mathbf{E}^{m}[n,j])}}
\end{equation}
For the $m$th head, the aggregated output of the node $v_n$ is given by $\mathbf{h}_n^m$:
\begin{equation}
	\mathbf{h}_n^m = \sum_{i=1}^{\mathcal{N}(n)} \alpha_{ni}^{m} \mathbf{g}_{ri}^{m}
\end{equation}
We concatenate the aggregated features of all nodes $\mathbf{H}^m = [\mathbf{h}_{1}^m,\dots,\mathbf{h}_{n}^m,\ldots,\mathbf{h}_{N}^m]$. The output of the aggregation module is:
\begin{equation}
	\mathbf{H} = \mathrm{concat} [\mathbf{H}^{1},\dots,\mathbf{H}^{m}, \ldots ,\mathbf{H}^{M}]
\end{equation}

\begin{figure}[t]
	\centering
	\includegraphics[scale=0.5]{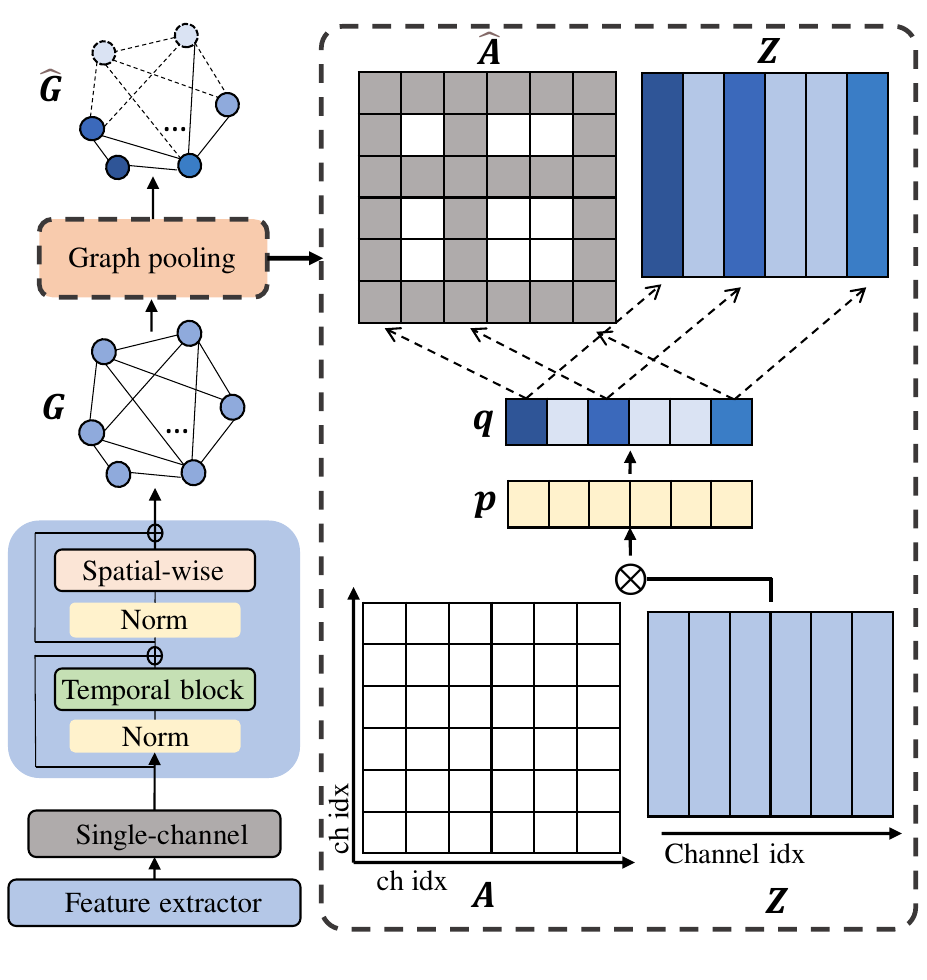}
	\caption{Graph-based channel selection based on graph pooling.}
	\label{fig:gpool}
\end{figure}

\section{Graph-based {channel} selection} \label{sec:channel_select}
 The graph-based channel selection algorithm $\mathscr{H}^{2}(\cdot)$ in \eqref{eq:channel_select} aims to select effective microphones from the spatial graph $G_{\mathrm{spatial}}^t$ by reconstructing the adjacent matrix $\mathbf{A}_{\mathrm{spatial}}$. In this section, we design two $\mathscr{H}^{2}(\cdot)$ according to whether the positions of the microphones or speakers are known.
	
\subsection{Graph-based channel selection with no prior information (gPool)}\label{subsec:gpool}
 When the positions of the microphones or speakers are unknown, we employ a graph-based pooling layer to select channels \cite{gao2019graph}. The procedure of the algorithm is in Fig.~\ref{fig:gpool}. Specifically, it first maps $\mathbf{Z}^t$ to a one-dimensional space using a learnable parameter $\mathbf{p} \in \mathbb{R}^{D}$:
\begin{equation}
	\mathbf{q} = \frac{\left(\mathbf{Z}^t\right)'\mathbf{p} }{ \|\mathbf{p}\|_2}
\end{equation}
where $\mathbf{q}\in\mathbb{R}^{C}$, $\|\cdot\|_2$ is the $\ell_2$-norm.
Then, it selects the channels that correspond to the top $K$ largest elements of $\mathbf{q}$ as the the most effective channels. Suppose the indices of the selected channels are:
\begin{equation}
	\mathrm{idx} = \mathrm{rank}(\mathbf{q},K)
\end{equation}
The reconstructed spatial graph is formulated as:
\begin{equation}
	\hat{\mathbf{Z}}^t[i,:] = \mathbf{Z}^{t}[i,\mathrm{idx}] \odot \left(\mathrm{sigmoid}(\mathbf{q}[\mathrm{idx}])\right)',\quad \forall i = 1,\ldots,D
\end{equation}
\begin{equation}
	\hat{\mathbf{A}}_{\mathrm{spatial}} = \mathbf{A}_{\mathrm{spatial}}[\mathrm{idx},\mathrm{idx}]
\end{equation}
where the symbol ``$:$'' means that all elements at the row or column of the matrix are selected. Finally, we have:
\begin{equation}
	\hat{\mathbf{Z}}_{\mathrm{CT}}= \{\hat{\mathbf{Z}}^{1},\dots,\hat{\mathbf{Z}}^{t},\dots,\hat{\mathbf{Z}}^{T}\}
\end{equation}
	
\subsection{Graph-based channel selection with prior knowledge ($\mathbf{A}_{\mathrm{prior}}$)} \label{subsec:prior}

\begin{figure}[t]
		\centering
		\includegraphics[scale=0.45]{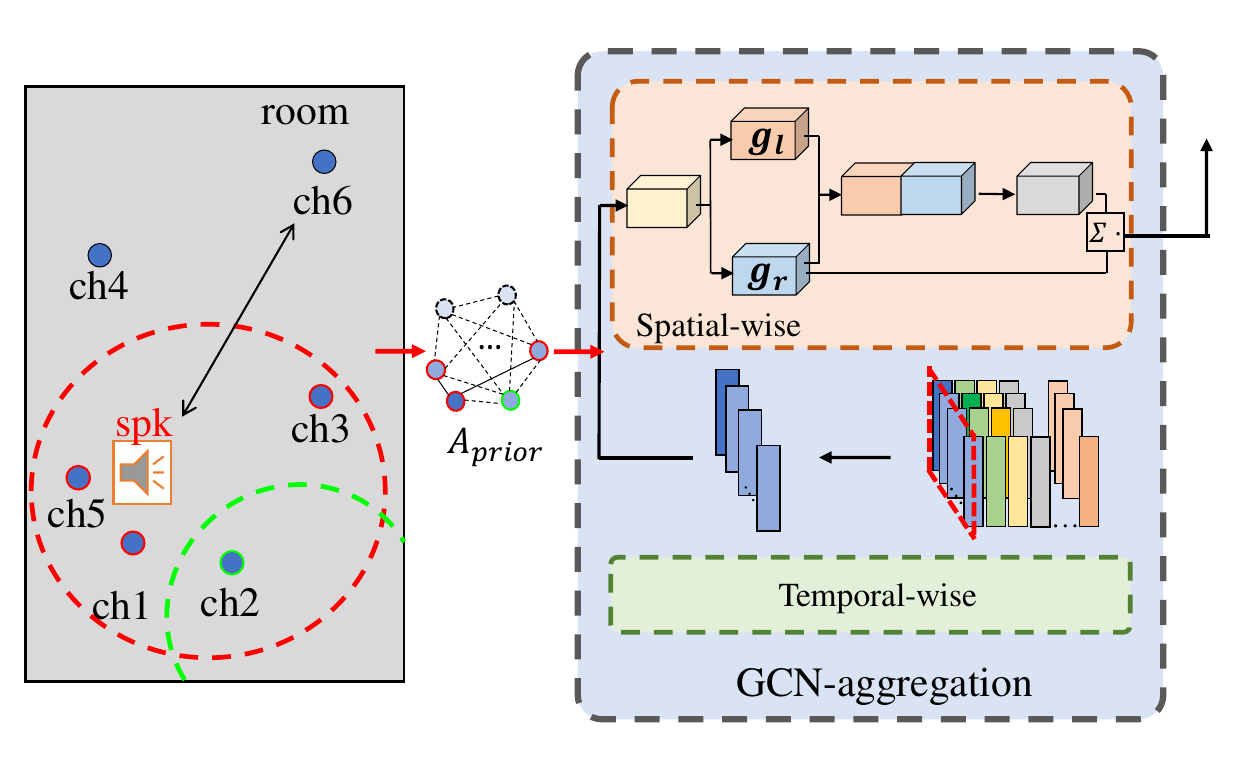}
		\caption{Graph-based channel selection with prior knowledge.}
		\label{fig:priorA}
\end{figure}

The direct estimation of the positions of the microphones and speakers is a hard issue. However, as shown in Fig. \ref{fig:priorA}, if the positions are known as a prior, then they may help the performance improvement significantly once properly utilized. In this situation, we rename $\mathbf{A}_{\mathrm{spatial}}$ as $\mathbf{A}_{\mathrm{prior}}$ to emphasize the utilization of the prior knowledge.

There are many ways to initialize $\mathbf{A}_{\mathrm{prior}}$ with different types of graphs in Section \ref{sec:relatedwork}. Here we choose the simple undirect graph, which initializes $\mathbf{A}_{\mathrm{prior}}$ with the relative distances between the microphones and speakers.
	
Specifically, we denote the distance between the speaker position and the $i$th microphone as $D_{\mathrm{(\mathrm{i},\mathrm{spk})}}$. The maximum of the distances is denoted as $D_{\mathrm{max}}$. Then, we construct $\mathbf{A}_{\mathrm{prior}} $ using a pre-defined scalar $\rho$ to choose the $K$-nearest channels:
	\begin{equation}
		\mathbf{A}_{\mathrm{prior}}[i,:] =
\left\{\begin{array}{ll}
1 ,&\quad \mathrm{if} \quad \frac{D_{(\mathrm{i},\mathrm{spk})}}{D_{\mathrm{max}}} < \rho \\
0,&\quad \mathrm{otherwise}
\end{array}
\right.
	\end{equation}
and set
	\begin{equation}
	\hat{\mathbf{Z}}_{\mathrm{CT}} = \mathbf{Z}_{\mathrm{CT}}
	\end{equation}

	Note that, if more abundant prior information is available, the adjacent matrix can be constructed in a more multiplicative way. For example, we can also utilize the location of the noise sources or the speaker orientation as a mask, which is further discussed in the ablation study in Section \ref{subsubsec: diifer_g}.
	
\section{Experimental setup} \label{sec:exp_setup}
	This section presents the experimental datasets as well as the parameter settings.
	
	
	\subsection{Datasets}
	
 The experiments were conducted on two simulation datasets---LibriSIMU-noise and LibriSIMU-reverb, as well as two real multi-channel datasets---Libri-adhoc40\cite{guan2021libri} and Hi-mia\cite{qin2020hi}. The detailed settings of the datasets are listed as follows.
	
	{\bf{LibriSIMU-noise}} is used to simulate the working scenario of speaker verification in noisy environment. For each utterance, we simulated a room environment. The width, length, and height of the room were selected randomly from ranges of $[8,10]$, $[12,14]$, and $[3,5]$ meters respectively. The reverberation environment was generated by the image-source library\footnote{https://github.com/DavidDiazGuerra/gpuRIR}. The reverberation time $T_{60}$ was selected randomly from a range of $[0.2,0.5]$ seconds. A single point source of noise, a single speaker, and an ad-hoc microphone array with 40 ad-hoc nodes were put randomly in the three dimensional space of the room, which implies a harsh experimental conditions. The speech source was from the train-clean-100, test, and dev subsets of Librispeech \cite{panayotov2015librispeech}. The noise data was from a mixed noise dataset \cite{tan2021speech} which includes classical bus noise, street noise, white noise, etc. The signal-to-noise ratio was in a range of $[-5,20]$ dB.

{\bf{LibriSIMU-reverb}} is used to simulate the working scenario of speaker verification in strong reverberation environment. The simulation is similar to the LibriSIMU-noise dataset, except that the reverberation time $T_{60}$ was in a range of $[0.2,1.2]$ seconds, and no additive noise was  added.

{\bf{Libri-adhoc40}}\cite{guan2021libri}: It is a replayed version of the Librispeech corpus in a real office environment.
The recording environment is an office room with a size of $9.8\times 10.3 \times 4.2$ meters. The room is highly reverberant with $T_{60}$ around 0.9 second and little additive noise. Each replayed utterance was recorded by an ad-hoc microphone array of 40 ad-hoc nodes. The locations of both the microphones and speakers of the training, evaluation, and test data are different. The distances between the speakers and the microphones were ranged from $0.8$ meter to $7.4$ meters, which makes the dataset suitable for the study of far-field speech processing.

	{\bf{Hi-mia}}\cite{qin2020hi}: It is a real text-dependent dataset for smart homes. It uses one close-talking microphone and six 16-channel microphone arrays to collect speech data. The training set takes AIshell-wakeup\footnote{https://www.aishelltech.com/wakeup\_data} as the speech source, which contains 254 speakers. The test set takes AIshell-2019B-eval\footnote{https://www.aishelltech.com/aishell\_2019\_eval} as the speech source, which contains 44 speakers. The text content of the speech source is 'ni hao mi ya' in Chinese and 'Hi Mia' in English with different speaking speed. In \cite{qin2020hi}, a speaker verification system was first trained with the text-independent single-channel AIshell-2\footnote{https://www.aishelltech.com/aishell\_2} data and data augmentation, and then fine-tuned with the text-dependent multichannel Hi-mia data. We follow the same step in this paper.

	\subsection{Parameter settings}
 The proposed method was implemented via the \emph{voxcelecb\_trainer} toolbox\footnote{{https://github.com/DavidDiazGuerra/gpuRIR}}.  {We stacked two spatial-temporal blocks.} The number of the attention head is 4.

  In the first training stage, the single-channel speaker verification system uses the architecture in \cite{chung2020defence}. It was pre-trained with Librispeech\cite{panayotov2015librispeech} train-clean-100 subset, when the test sets of the LibriSIMU-noise, LibriSIMU-reverb, and Libri-adhoc40 dataset were used as the test data. It was pre-trained with the \emph{iOS} subset of AIShell-2\cite{du2018aishell}, when the test set of the Hi-mia dataset was used as the test data. The best model among the 200 training epochs of the single-channel system was used to initialize the multichannel system for the second-stage training. In the second training stage, we randomly selected 20 channels for each training utterance as the multichannel training data. For each utterance of a test set, we randomly selected $\{8, 16, 32, 40\}$ channels respectively for evaluation.

Multiple variants of the proposed method were used for evaluation, which are denoted as follows:
\begin{itemize}
	\item\textbf{Self-attention aggregation with graphs as masks (SAM-agg)}: {It only contains the SAM-agg spatial-temporal aggregation block. No channel selection block is added.}
	
	\item\textbf{Graph convolution network based aggregation (GCN-agg)}: {It only contains the GCN-agg spatial-temporal aggregation block. No channel selection block is added.}
	
	\item \textbf{Channel selection based on graph pooling with GCN-agg mechnism (GCN-agg+gpool)}: {It adds the gPool channel selection block after the GCN-agg block.}
	
	\item \textbf{Channel selection based on prior adjacent matrix with GCN-agg mechnism (GCN-agg+$\mathbf{A}_{\mathrm{\mathbf{prior}}}$)}: {It adds the $\mathbf{A}_{\mathrm{prior}}$ channel selection block after the GCN-agg block.}
\end{itemize}

We used the equal error rate (EER) as the evaluation metric, and reported the number of parameters of the comparison methods.

	\begin{table*}[t]
		\centering
		\caption{EER (\%) of the comparison methods on simulated data. The test scenario ``$N$-channels'' indicates that an ad-hoc microphone array contains $N$ ad-hoc nodes, each of which consists of a single microphone.}
		\def \temptablewidth{0.9\textwidth}
		\begin{tabular*}{\temptablewidth}{@{\extracolsep{\fill}}cccccccc}
			\toprule[1.3pt]
			Dataset & Description & Algorithm & Model size & 8-channels & 16-channels & 32-channels & 40-channels \\
			\hline
			
			\multirow{9}{*}{LibriSIMU-noise} & \multirow{2}{*}{Single channel} &  Oracle One-best & 1.437 M & 35.848 & 34.7498 & 33.3556 & 33.7853 \\
			\cline{3-8}
			& & Beamforming  & 1.437 M &  37.3663 & 36.6119 &  32.1176 & 31.5578 \\
			\cline{2-8}
			
			& \multirow{3}{*}{Multi-channel} & MEAN-uttr-agg & 1.437 M & 38.4645 & 37.7483 & 38.2066  & 37.7483 \\
			\cline{3-8}
			& & MHA-uttr-agg & 1.503 M & 22.8132 & 21.2567 & 19.6047 & 19.9771 \\
			\cline{3-8}
			& & AP-uttr-agg & 1.454 M & 22.8610 & 21.6100 & 20.3495 & 20.7792 \\
			\cline{2-8}
			&\multirow{4}{*}{\textbf{Proposed}} & SAM-agg & 1.570 M & 21.0657 & 19.8830 & 18.1341  & 18.5447 \\
			\cline{3-8}
			& & GCN-agg & 1.502 M   & 22.2116 & 20.3972 & 19.5187 & 19.4041  \\
			\cline{3-8}
			&  & GCN-agg+gPool & 1.502 M   & 22.0015 & 21.0752 & 19.5092  & 19.6524 \\
			\cline{3-8}
			& & \textbf{GCN-agg+$\mathbf{A}_{\mathbf{prior}}$} & 1.502 M  & \textbf{18.7353} & \textbf{17.5890} & \textbf{16.5871} & \textbf{16.6062} \\
			
			\hline
			
			\multirow{9}{*}{LibriSIMU-reverb} & \multirow{2}{*}{Single channel} &  Oracle one-best & 1.437 M & 33.6612 & 33.0405 & 32.4120  & 32.4389 \\
			\cline{3-8}
			& & EV & 1.437 M & 33.7853 & 32.9736 & 32.2765  & 32.4771\\
			\cline{2-8}
			
			& \multirow{3}{*}{Multi-channel} & MEAN-uttr-agg & 1.437 M & 32.7540 & 32.2956 & 32.1906  & 32.6299 \\
			\cline{3-8}
			& & MHA-uttr-agg & 1.503 M & 19.0126 & 17.924 & 17.1028  & 17.0646 \\
			\cline{3-8}
			& & AP-uttr-agg & 1.454 M & 22.8610 & 21.6192 & 20.3495  & 20.7792 \\
			\cline{2-8}
			&\multirow{4}{*}{\textbf{Proposed}} & SAM-agg & 1.570 M & 17.5882 & 16.4725 & 16.0332 & 15.8804 \\
			\cline{3-8}
			& & GCN-agg & 1.502 M  & 19.1176 & 18.2391 & 17.7617  & 17.8285 \\
			\cline{3-8}
			&  & GCN-agg+gPool & 1.502 M & 18.7166 & 17.5802 & 17.1791  & 17.1982 \\
			\cline{3-8}
			& & \textbf{GCN-agg+$\mathbf{A}_{\mathbf{prior}}$} & 1.502 M & \textbf{15.2884}  & \textbf{13.7987} &  \textbf{13.9324}  & \textbf{13.8178} \\
			\toprule[1.3pt]
		\end{tabular*}
		\label{table:main_simu}
	\end{table*}

	\begin{table*}[t]
		\centering
		\caption{EER (\%) of the comparison methods on real-world datasets.}
		\def \temptablewidth{0.9\textwidth}
		\begin{tabular*}{\temptablewidth}{@{\extracolsep{\fill}}cccccccc}
			\toprule[1.3pt]
			Dataset & Description & Algorithm & Model size & 8-channels & 16-channels & 32-channels & 40-channels \\
			\hline
			
			\multirow{10}{*}{Hi-mia} & \multirow{3}{*}{Single channel} &  Oracle one-best & 1.437 M & 10.4262 & 10.3774 & 10.1171  & 10.4831 \\
			\cline{3-8}
			& & EV & 1.437 M & 11.6054 & 11.0849 & 11.1093 & 11.2069 \\
			\cline{3-8}
			& & Beamforming & 1.437 M &  22.7197 & 23.3874 & 19.5373  & 20.4728\\
			\cline{2-8}
			
			& \multirow{3}{*}{Multi-channel} & MEAN-uttr-agg & 1.437 M & 10.7921 & 10.8328 & 9.2306  & 9.3770 \\
			\cline{3-8}
			& & MHA-uttr-agg & 1.503 M & 8.7996 & 8.7833 & 7.7586 & 8.2059 \\
			\cline{3-8}
			& & AP-uttr-agg & 1.454 M & 8.5068 & 8.0433 & 7.4170  & 7.4252 \\
			\cline{2-8}
			&\multirow{4}{*}{\textbf{Proposed}} & SAM-agg & 1.507 M & 8.9460 & 8.6126 & 7.9375  & 8.0270 \\
			\cline{3-8}
			& & \textbf{GCN-agg}  & 1.502 M  & \textbf{7.6610} & \textbf{7.6122} & \textbf{7.2137}  & \textbf{7.2869} \\
			\cline{3-8}
			&  & GCN-agg+gPool & 1.502 M & 7.9213  & 7.9213 &  7.3276  & 7.4252 \\
			\cline{3-8}
			& & GCN-agg+$\mathbf{A}_{\mathrm{prior}}$ & -  & - & - & - & - \\
			
			\hline
			
			\multirow{10}{*}{Libri-adhoc40} & \multirow{3}{*}{Single channel} &  Oracle one-best & 1.437 M & 19.5420 & 21.4981 & 19.3130  & 18.3111\\
			\cline{3-8}
			& & EV & 1.437 M & 19.6103 & 20.8206 & 19.3225  & 19.5802\\
			\cline{3-8}
			& & Beamforming  & 1.437 M &  21.3836 & 21.4408 & 16.6698  & 16.3168\\
			\cline{2-8}
			
			& \multirow{3}{*}{Multi-channel} & MEAN-uttr-agg & 1.437 M & 18.3015 & 18.1679 & 18.0821  & 17.9294 \\
			\cline{3-8}
			& & MHA-uttr-agg & 1.503 M & 10.7061 & 11.0496 & 10.0477  & 10.2767 \\
			\cline{3-8}
			& & AP-uttr-agg & 1.454 M & 11.1450 & 11.4218 & 11.1450 & 11.0782 \\
			\cline{2-8}
			&\multirow{4}{*}{\textbf{Proposed}} & SAM-agg & 1.570 M & 9.5802 & 9.8282 & 9.0458 & 9.2557 \\
			\cline{3-8}
			& & GCN-agg & 1.502 M & 11.0687 & 10.4389 & 9.9141  & 9.8760  \\
			\cline{3-8}
			&  & GCN-agg+gPool & 1.502 M & 11.2023 & 10.7634 & 9.7901  & 9.4179 \\
			\cline{3-8}
			& & \textbf{\textbf{GCN-agg+}$\mathbf{A}_{\mathbf{prior}}$} & 1.502 M &
			\textbf{8.5878} & \textbf{9.0935} & \textbf{7.8626} & \textbf{8.0057} \\
			
			\toprule[1.3pt]
		\end{tabular*}
		\label{table:main_real}
	\end{table*}
	
\subsection{Comparison methods}
	
We compared with 6 referenced methods, which can be categorized into the following two classes. The first class aims to generate a single channel speech signal from the multiple channels of an ad-hoc microphone array, and then applies it to a single-channel speaker verification system:
	\begin{itemize}
\item \textbf{Oracle one-best}: It selects a physically closest channel to the speech source.
	
\item  \textbf{Beamforming} \cite{anguera2007acoustic}: It uses a conventional delay-and-sum beamforming to aggregate all channels into a single channel.
	
\item  \textbf{EV} \cite{wolf2014channel}: It selects a channel whose energy envelope has the highest variance among all channels.
\end{itemize}

The second class uses a multichannel speaker verification system to handle the speech signals from ad-hoc microphone arrays directly:
	\begin{itemize}
\item\textbf{Utterance-level channel mean aggregation (MEAN-uttr-agg)}: It extracts an utterance-level speaker embedding from each channel, and then averages the speaker embeddings of all channels into a single-channel speaker embedding.
	
\item\textbf{Utterance-level channel aggregation based on multihead attention (MHA-uttr-agg)} \cite{liang2022apsipa}: It extracts an utterance-level speaker embedding from each channel, and then aggregates the speaker embeddings using the  multi-head attention mechanism.
	
\item \textbf{Utterance-level channel aggregation based on attentive pooling (AP-uttr-agg)}\cite{cai2021embedding}: It extracts an utterance-level speaker embedding from each channel, and then conducts the weighted average over the speaker embeddings of all channels where the weights of the channels are calculated by an attention pooling layer.
	\end{itemize}

\section{{Experimental results}} \label{sec:result}

In this section, we first show the main comparison results in Section \ref{subsec:main}, then show the robustness of the proposed method to the variation of the noise and reverberation environments in Section \ref{subsec:simu}, and finally study the effects of the components of the proposed method on performance in Sections \ref{subsubsec: diifer_g} and \ref{subsec:ab}.

\subsection{Main results}\label{subsec:main}

Table~\ref{table:main_simu} lists the comparison results on the simulated datasets. From the table, we see that, the proposed graph-based methods outperform all referenced methods. {We take the 8-channels test scenario as an example.} SAM-agg achieves a relative EER reduction of $7.66\%$ over MHA-uttr-agg, and $ 7.64\%$ over the AP-uttr-agg. Even the GCN-agg, which performs the poorest among the variants of the proposed method, outperforms the best utterance-level method AP-uttr-agg by a relative EER reduction of $2.71\%$. A similar phenomenon is observed on the libriSIMU-reverb dataset as well.

Comparing the variants of the proposed methods, we observe the following phenomena. GCN-agg+gpool obtains $2.09\%$ relative EER reduction over GCN-agg on libriSIMU-reverb. However, this advantage was not transferred to libriSIMU-noise. GCN-agg+$\mathrm{A}_{\mathrm{prior}}$  outperforms GCN-agg by a relative EER reduction of  $15.65\%$ on LibriSIMU-noise, and $20.02\%$ on LibriSIMU-reverb. Moreover, it achieves a relative EER reduction of $12.99\%$ over the runner-up method SAM-agg on LibriSIMU-reverb, and $6.40\%$ on LibriSIMU-noise, which demonstrates the importance of prior knowledge in the study of ad-hoc microphone arrays.

Table~\ref{table:main_real} lists the comparison results on the real datasets. From the table, we see that the proposed methods significantly outperform the referenced methods, which is similar to that on the simulated test data. Comparing the variants of the proposed methods, we further  observe the following phenomena. GCN-agg outperforms SAM-agg on Hi-mia, e.g. by a relative EER reduction of $14.36\%$ on the 8-channels test scenario. GCN-agg+gPool outperforms  GCN-agg by at most a relative EER reduction of $4.64\%$ on Libri-adhoc40, while GCN-agg+$\mathrm{A}_{\mathrm{prior}}$ obtains a relative EER reduction of $18.94\%$ over GCN-agg, on the 8-channels test scenario.

\begin{figure*}[t]
	\centering
	\includegraphics[scale=0.5]{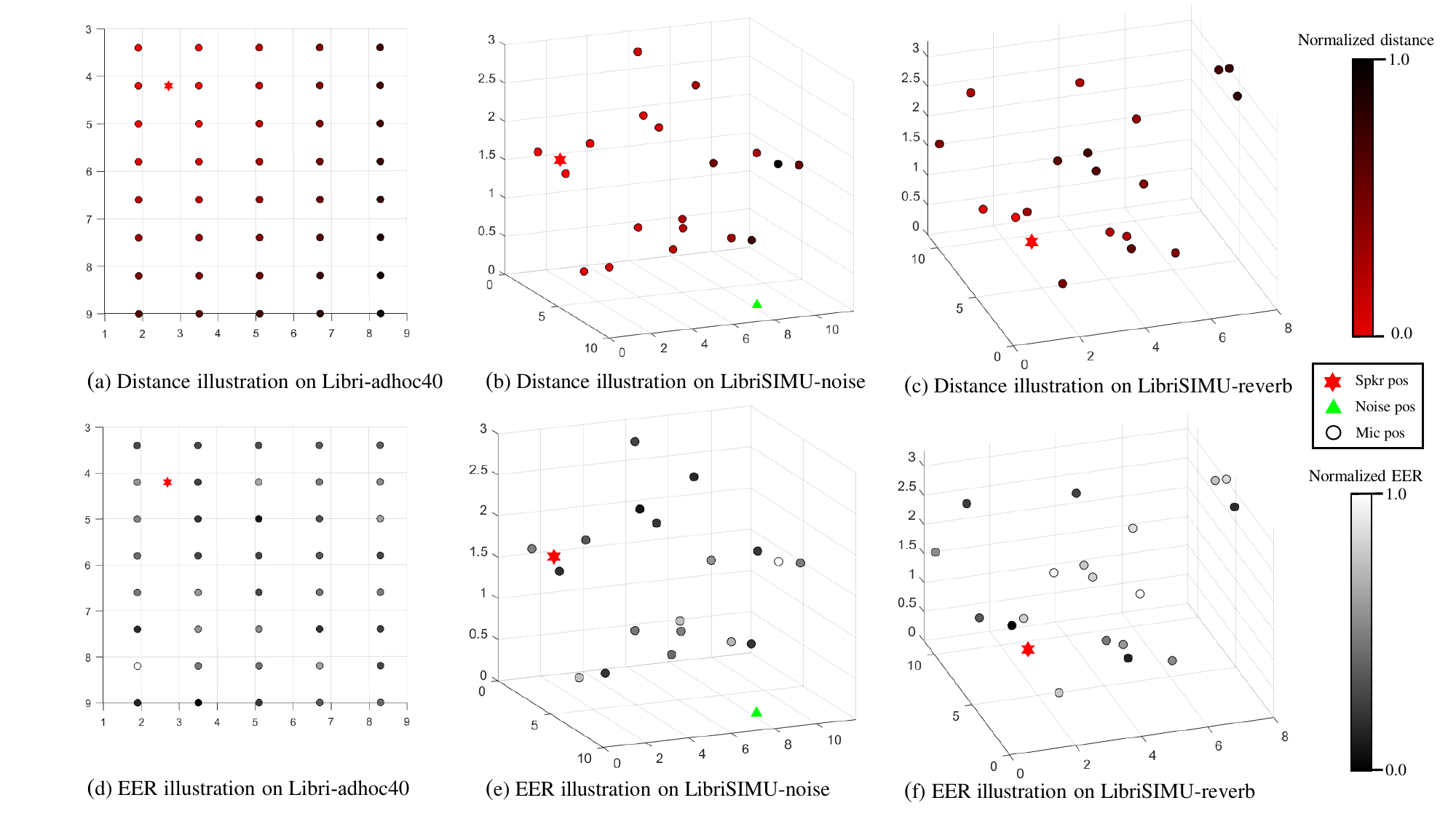}
	\caption{EER performance of speaker verification at each node of an ad-hoc microphone array with respect to the distance between the node and a speaker, where each node contains a single microphone. Figs.~\ref{fig:visualize_exp}a to \ref{fig:visualize_exp}c show the relative distance between the microphone nodes and the speaker. Figs.~\ref{fig:visualize_exp}d to \ref{fig:visualize_exp}f show the EER performance at the nodes. The three dimensions in a subfigure indicates the length, width, and height of a room respectively. }
	\label{fig:visualize_exp}
\end{figure*}

Fig. \ref{fig:visualize_exp} analyzes the performance of speaker verification at each node of an ad-hoc microphone array with respect to the distance between the node and a speaker, where the model GCN-agg is used as the speaker verification system. From the figure, we see that the EER of the microphone nodes is correlated with the distance from the microphones to the speaker and point noise source. For example, comparing Fig.~\ref{fig:visualize_exp}a with Fig.~\ref{fig:visualize_exp}d, we see that the channels close to the speaker yield good performance on the Libri-adhoc40. Similar phenomenon is observed on the LibriSIMU-reverb as well. Comparing Fig.~\ref{fig:visualize_exp}b with Fig.~\ref{fig:visualize_exp}e, we see that most of the channels that yield good performance are not only close to the speaker, but also far away from the point noise source. However, an important phenomenon is that some channels that are close to the wall also yield excellent performance, though they are far away from the speaker. At last, we observe that GCN-agg is able to grasp the spatial-temporal difference between the channels.

\begin{table}[t]
	\centering
	\caption{EER (\%) of the comparison methods on the 20-channel test scenarios of the LibriSIMU-noise simulated data with different SNR levels.}
	\def \temptablewidth{0.6\textwidth}
	\scalebox{0.8}{
	\begin{tabular*}{\temptablewidth}{@{\extracolsep{\fill}}cccccc}
		\toprule[1.3pt]
		\multirow{3}{*}{\diagbox{Algorithm}{SNR (dB)}} & \multirow{3}{*}{[-5,0]} & \multirow{3}{*}{[0,5]} & \multirow{3}{*}{[5,10]} & \multirow{3}{*}{[10,15]} & \multirow{3}{*}{[15,20]} \\
		&  & &  & &  \\
		&  & &  & & \\
		\hline
		Orcale one-best& 40.6990 & 32.7158 & 28.7051 & 27.0340 & 25.3915 \\
		\hline
		beamforming & 48.6631  & 39.3239 & 27.3014 & 25.1337 & 22.9183 \\
		\hline
		MEAN-uttr-agg & 42.9240 & 32.6967 & 27.2981 & 27.2823 & 25.3915 \\
		\hline
		MHA-uttr-agg & 28.7147 & 20.1585 & 16.5011 & 16.4057 & 16.8311 \\
		\hline
		AP-uttr-agg & 29.6123 & 20.3209 & 16.9118 & 17.0168 & 17.5229 \\
		\hline
		SAM-agg & 25.5971 & 17.0794 & 15.4698 & 15.2311 & 15.1547 \\
		\hline
		GCN-agg & 28.9152 & 19.6142 & 16.6921 & 16.1860 & 15.8804 \\
		\hline
		GCN-agg+gPool & 28.1895 & 20.1776 & 16.8927 & 16.7876 & 16.2911 \\
		\hline
		\textbf{GCN-agg+$\textbf{A}_{\textbf{prior}}$} & \textbf{25.5252} & \textbf{16.8258} & \textbf{13.7319} & \textbf{12.7383} & \textbf{13.0252} \\
		\toprule[1.3pt]
	\end{tabular*}}
	\label{table:robust}
\end{table}

\begin{table}[t]
	\centering
	\caption{EER (\%) of the comparison methods on the 20-channel test scenarios of the LibriSIMU-reverb simulated data with different reverberation time.}
	\def \temptablewidth{0.62\textwidth}
	\scalebox{0.8}{
	\begin{tabular*}{\temptablewidth}{@{\extracolsep{\fill}}cccccc}
		\toprule[1.3pt]
		\multirow{3}{*}{\diagbox{Algorithm}{$T_{60}$ (seconds)}} & \multirow{3}{*}{$[0.2,0.4)$} & \multirow{3}{*}{$[0.4,0.6)$} & \multirow{3}{*}{$[0.6,0.8)$} & \multirow{3}{*}{$[0.8,1.0)$} & \multirow{3}{*}{$[1.0,1.2]$}\\
		&  & &  & & \\
		&  & &  & & \\
		\hline
		Oracle one-best& 23.2784 & 28.4091 & 31.9519 & 35.2941 & 36.2303 \\
		\hline
		EV & 23.3065 & 28.3327 & 32.8182 & 35.2464 & 35.8636 \\
		\hline
		MEAN-uttr-agg &25.5614 & 28.3040 & 31.8277 & 34.0909 & 34.9743 \\
		\hline
		MHA-uttr-agg & 14.5865 & 15.7086 & 17.1982 & 18.7643 & 20.297 \\
		\hline
		AP-uttr-agg & 15.5782 & 17.1123 & 18.5733 & 20.4545 & 21.5255 \\
		\hline
		SAM-agg & 13.6321 & 14.9351 & 15.8231 & 17.4656 & 18.0986 \\
		\hline
		GCN-agg & 15.5034 & 16.8354 & 17.9049 & 19.7383 & 20.2145 \\
		\hline
		GCN-agg+gPool &14.5397 & 16.3293 & 17.1409 & 19.0317 & 19.9487 \\
		\hline
		\textbf{GCN-agg+$\textbf{A}_{\textbf{prior}}$} & \textbf{12.0790} & \textbf{12.5286} & \textbf{13.8369} & \textbf{15.3266} & \textbf{16.8500} \\
		\toprule[1.3pt]
	\end{tabular*}}
	\label{table:robust2}
\end{table}

\subsection{Results in different noise and reverberation conditions}\label{subsec:simu}

To study how the proposed method behaves in different noise and reverberation conditions, we preset the SNR and reverberation time $T_{60}$ of the test scenarios. Table~\ref{table:robust} lists the EER of the comparison methods on the 20-channels test scenario of LibriSIMU-noise with different SNR levels. From the table, we see that, although the performance of all comparison methods is getting worse when the test scenarios become more challenge, the proposed methods outperform the referenced methods, and GCN-agg+$\mathrm{A}_{\mathrm{prior}}$ performs the best in all cases. Specifically, when $\mathrm{SNR} \in [-5,0]\mathrm{dB}$, SAM-agg obtains a relative EER reduction of $10.86\%$ over the best referenced method MHA-uttr-agg. When $\mathrm{SNR}$ is enlarged to, e.g., $[15,20]\mathrm{dB}$, the relative EER reduction is still $9.96\%$.
 GCN-agg+$\mathrm{A}_{\mathrm{prior}}$ achieves a relative EER reduction of $11.0\%$ over SAM-agg.

Table~\ref{table:robust2} lists the EER of the comparison methods on the 20-channels test scenario of LibriSIMU-reverb with different reverberation time. The experimental phenomena are similar with those in Table \ref{table:robust}. Specifically, when the reverberation time $T_{60} \in [0.2,0.4]$ seconds, SAM-agg obtains a relative EER reduction of $6.53\%$ over the best referenced method MHA-uttr-agg. GCN-agg+$A_{\mathrm{prior}}$ further reduces EER by relatively $6.90\%$ over SAM-agg. When $T_{60} \in [1.0,1.2]$ seconds, SAM-agg obtains a relative EER reduction of $10.83\%$ over MHA-uttr-agg. GCN-agg+$\mathrm{A}_{\mathrm{prior}}$ further outperforms SAM-agg by a relative EER reduction of $11.39\%$.

\subsection{Effects of the adjacent matrices on performance} \label{subsubsec: diifer_g}
	
The adjacent matrices $\mathbf{A}_{\mathrm{temporal}}$ and $\mathbf{A}_{\mathrm{spatial}}$ can be constructed in different ways, where $\mathbf{A}_{\mathrm{spatial}}$ is rewritten as $\mathbf{A}_{\mathrm{prior}}$ when it is constructed with the spatial information prior. In this section, we study how the construction methods affect the performance. Particularly, we denote $\mathbf{A}_{\mathrm{prior}}$ with the prior knowledge ``$X$'' as $\mathbf{A}_{\mathrm{prior}}^{X}$.

Table~\ref{table:ab_g} lists the effect of the construction methods of the adjacent matrices on the Libri-adhoc40 dataset. Specifically, $\mathbf{A}_{\mathrm{temporal}}$ is constructed by the two ways in Section \ref{subsec:temporal} where the parameter $\delta$ in $\mathrm{span}(j,\delta)$ is set to $\{0,1\}$. From the table, we see that constructing $\mathbf{A}_{\mathrm{temporal}}$  with the complete graph is better than that with the sparse graph $\mathrm{span}(j,\delta)$.

\begin{table}[t]
	\centering
	\caption{EER (\%) of the proposed multi-channel speaker verification with different construction methods of the adjacent matrices on the Libri-adhoc40 test set.}
	\def \temptablewidth{0.6\textwidth}
	\scalebox{0.8}{
	\begin{tabular*}{\temptablewidth}{@{\extracolsep{\fill}}cccccc}
		\toprule[1.3pt]
	Name & Setting & 8-channels & 16-channels & 32-channels & 40-channels \\
		\hline
		\multirow{3}{*}{ $\mathbf{A}_{\mathrm{temporal}}$  } & $\mathrm{span}(j,0)$ & 11.3263 & 11.7366 & 11.3263 & 11.4408 \\
		\cline{2-6}
		& $\mathrm{span}(j,1)$ & 10.5916 & 11.1164 & 10.6679 & 10.687 \\
		\cline{2-6}
		& complete & 11.0687 & 10.4389 & 9.9141 & 9.8760 \\
		\hline
		
		\multirow{4}{*}{ $\mathbf{A}_{\mathrm{prior}}^{\mathrm{pos}}$ } & $\rho=0.1$ & 8.9313 & \textbf{8.6069} & 8.1489 & 8.3779 \\
		\cline{2-6}
		& $\rho=0.3$ & \textbf{7.9198} & 8.7500 & \textbf{7.6622} & \textbf{7.8626} \\
		\cline{2-6}
		& $\rho=0.6$ & 8.5878 & 9.0935 & 7.8626 & 8.0057 \\
		\cline{2-6}
		& $\rho=0.9$ & 8.9599 & 9.1508 & 8.3206 & 8.6641 \\
		\hline
		
		\multirow{4}{*}{ $\mathbf{A}_{\mathrm{prior}}^{\mathrm{pos+ori}}$ } & $\rho=0.1$ & 8.3492 & 8.8836 & 8.3588 & 8.8931 \\
		\cline{2-6}
		& $\rho=0.3$ & \textbf{7.0802} & \textbf{8.5495} & 8.2634 & 8.2443 \\
		\cline{2-6}
		& $\rho=0.6$ & 7.7481 & 8.5496 & 8.2601 & 8.3015 \\
		\cline{2-6}
		& $\rho=0.9$ & 8.6927 & 8.6641 & \textbf{7.6908} & \textbf{7.8912} \\
		\toprule[1.3pt]
	\end{tabular*}}
	\label{table:ab_g}
	\end{table}

\begin{table}[t]
		\centering
		\caption{Effect of the construction methods of the adjacent matrices on the LibriSIMU-noise test set in terms of EER (\%). }
		\def \temptablewidth{0.6\textwidth}
		\scalebox{0.8}{
		\begin{tabular*}{\temptablewidth}{@{\extracolsep{\fill}}cccccc}
			\toprule[1.3pt]
			\multirow{2}{*}{\diagbox{Algorithm}{SNR (dB)}} & \multirow{2}{*}{[-5,0]} & \multirow{2}{*}{[0,5]} & \multirow{2}{*}{[5,10]} & \multirow{2}{*}{[10,15]} & \multirow{2}{*}{[15,20]} \\
			&  & &  & &  \\
			\hline
			$\mathbf{A}_{\mathrm{prior}}^{\mathrm{pos}}$ & 26.2987 & 17.3329 & 13.9037 & 13.3403 & 13.0443 \\
			\hline
			$\mathbf{A}_{\mathrm{prior}}^{\mathrm{pos+noise\_pos}}$  & \textbf{25.5252} & \textbf{16.8258} & \textbf{13.7319} & \textbf{12.7387} & \textbf{13.0252} \\
			\toprule[1.3pt]
		\end{tabular*}}
		\label{table:ab_simu}
	\end{table}

\begin{table}[t]
		\centering
		\caption{EER (\%) of the multichannel speaker verification with different adjacent matrices on the LibriSIMU-reverb test set. }
		\def \temptablewidth{0.7\textwidth}
		\scalebox{0.7}{
		\begin{tabular*}{\temptablewidth}{@{\extracolsep{\fill}}cccccc}
		\toprule[1.3pt]
			\multirow{2}{*}{\diagbox{Algorithm}{$T_{60}$ (seconds)}} & \multirow{2}{*}{-} & \multirow{2}{*}{8-channels} & \multirow{2}{*}{16-channels} & \multirow{2}{*}{32-channels} & \multirow{2}{*}{40-channels}\\
			&  & &  & & \\
			\hline
			$\mathbf{A}_{\mathrm{prior}}^{\mathrm{pos}}$ & - & 18.9553 & 17.9530 & 17.1696 & 17.0932 \\
			\hline
			$\mathbf{A}_{\mathrm{prior}}^{\mathrm{pos+noise\_pos}}$  & - & \textbf{18.7353} & \textbf{17.5898} & \textbf{16.5871} & \textbf{16.6062} \\
		\toprule[1.3pt]
	\end{tabular*}}
		\label{table:ab_simu2}
	\end{table}

\begin{table*}[t!]
		\centering
		\caption{Ablation study of the temporal and spatial modules of the proposed method on real-world data in terms of EER (\%).}
		\def \temptablewidth{0.8\textwidth}
		\scalebox{0.9}{
		\begin{tabular*}{\temptablewidth}{@{\extracolsep{\fill}}cccccc}
		\toprule[1.3pt]
 		Dataset & Algorithm & 8-channels & 16-channels & 32-channels & 40-channels \\
 		\hline
 		\multirow{4}{*}{Libri-adhoc40} & MEAN-uttr-agg & 18.3015 & 18.1679 & 18.0821 & 17.9294 \\
 		\cline{2-6}
 		& GCN-agg-temporal & 11.7176 & 11.6126 & 11.4790 & 11.8511  \\
 		\cline{2-6}
 		& GCN-agg-spatial & \textbf{10.8111} & 11.1355 & 10.4676 & 10.5439 \\
 		\cline{2-6}
 		& \textbf{GCN-agg} & 11.0687 & \textbf{10.4389} & \textbf{9.9141} & \textbf{9.8760} \\
		\hline
		\multirow{4}{*}{Hi-mia} & MEAN-uttr-agg & 10.7921 & 10.8328 & 9.2306 & 9.3770 \\
		\cline{2-6}
		 & GCN-agg-temporal & 8.7202 & 8.3215 & 8.1057 & 7.6986 \\
		\cline{2-6}
		& GCN-agg-spatial & 7.9619 & 7.6668 & 7.2381 & 7.3845 \\
		\cline{2-6}
		& \textbf{GCN-agg} & \textbf{7.6610} & \textbf{7.6122} & \textbf{7.2137} & \textbf{7.2869} \\
		\toprule[1.3pt]
		\end{tabular*}}
		\label{table:ab_s_t}
\end{table*}

$\mathbf{A}_{\mathrm{prior}}^{\mathrm{pos}}$ is constructed using the position information prior of the ad-hoc microphone arrays and the sound sources, as presented in Section~\ref{subsec:prior}, where the tunable parameter $\rho$ determines the number of selected channels. A small $\rho$ means that the selected channels are close to the speaker. As shown in Table~\ref{table:ab_g}, when $\rho$ decreases from $ 0.9 $ to $0.3$, the performance is improved; however, when $\rho$ is further reduced to $ 0.1$, the performance decreases due to that only very limited number of channels are selected.

$\mathbf{A}_{\mathrm{prior}}^{\mathrm{pos+ori}}$ uses the speaker orientation as an additional prior where the ad-hoc nodes that are placed behind the speaker are masked off by further setting the corresponding elements of $\mathbf{A}_{\mathrm{prior}}^{\mathrm{pos}}$ to zero. As shown in Table~\ref{table:ab_g}, when $\rho = 0.9$, the performance of $\mathbf{A}_{\mathrm{prior}}^{\mathrm{pos+ori}}$ on the 40-channel test scenario outperforms that produced by $\mathbf{A}_{\mathrm{prior}}^{\mathrm{pos}}$. However, when $\rho$ is gradually reduced, the advantage of  $\mathbf{A}_{\mathrm{prior}}$+mask over $\mathbf{A}_{\mathrm{prior}}$ is limited.

Tables~\ref{table:ab_simu} and \ref{table:ab_simu2} list the effect of the construction methods of the adjacent matrices on the LibriSIMU-noise and LibriSIMU-reverb datasets respectively, where $\mathbf{A}_{\mathrm{prior}}^{\mathrm{pos+noise\_pos}}$ denotes that the microphone nodes that are close to the point noise source are further masked off.  From the tables, we see that $\mathbf{A}_{\mathrm{prior}}^{\mathrm{pos+noise\_pos}}$  improves the performance in the low SNR scenarios, which obtains a relative EER reduction of $2.94\%$ over $\mathbf{A}_{\mathrm{prior}}$.

\subsection{Effect of the graph-based spatial-temporal aggregation
block on performance}\label{subsec:ab}

The graph-based spatial-temporal aggregation block consists of two components---a temporal module and a spatial module. In this section, we analyze the effects of the temporal module and spatial module respectively.

Specifically, we compare the graph-based temporal module with a method of simply averaging the features along the time dimension, i.e. ``mean pooling over time'', and compare the graph-based  spatial module with a method of averaging the features along the spatial dimension, i.e..``mean pooling over space'', which derives the following three comparison methods:
\begin{itemize}
  \item \textbf{MEAN-uttr-agg}: Both the graph-based temporal and spatial modules of the proposed method are replaced by their corresponding mean pooling strategies.
  \item \textbf{GCN-agg-temporal}: The graph-based spatial module is replaced by the mean pooling over space.
  \item \textbf{GCN-agg-spatial}: The graph-based temporal module is replaced by mean pooling over time.
\end{itemize}

Table \ref{table:ab_s_t} lists the EER performance of the GCN-agg variants on the real-world data. From the table, we see that the performance of the proposed method performs the best; compared with ``MEAN-uttr-agg'', the performance improvement produced by GCN-agg-spatial is more significant than that produced by ``GCN-agg-temporal''. We take the 8-channels test scenario as an example. ``GCN-agg-spatial'' outperforms ``MEAN-uttr-agg'' by a relative EER reduction of $40.93\%$ on Libri-adhoc40 and $26.22\%$ on Hi-mia respectively, while ``GCN-agg-temporal'' outperforms ``MEAN-uttr-agg'' by a relative EER reduction of only $35.97\%$ on Libri-adhoc40 and $19.20\%$ on Hi-mia respectively.

\section{Conclusions} \label{sec:conclusion}
	In this paper, we have proposed the graph-based frame-level multi-channel speaker verification system with ad-hoc microphone arrays. It consists of two components---graph-based spatial-temporal aggregation block and graph-based channel selection block. The spatial-temporal aggregation block first uses graphs to model the interdependencies of frame-level multichannel speaker embeddings, then aggregates the embeddings along both the temporal and spatial dimensions. The channel selection block further chooses channels that are most helpful to the performance improvement. The core novelties lie in that, to our knowledge, the proposed method models multichannel speaker verification with graphs for the first time; moreover, it constructs adjacent matrices of the graphs with environmental priors flexibly. We have compared the proposed method with a number of representative algorithms on both simulated and real-world datasets. Experimental results in both scenarios show that the proposed method significantly outperforms the referenced methods. For example, the proposed method outperforms the best referenced method by a relative EER of at least $15.39\%$ in the simulated noisy environment, and $18.54\%$ in the simulated reverberant environment. Moreover, the proposed method is robust against the variation of reverberation and SNR levels.
	
	\bibliographystyle{IEEEtran}
	\bibliography{Reference}

\begin{thebibliography}{10}
\providecommand{\url}[1]{#1}
\csname url@samestyle\endcsname
\providecommand{\newblock}{\relax}
\providecommand{\bibinfo}[2]{#2}
\providecommand{\BIBentrySTDinterwordspacing}{\spaceskip=0pt\relax}
\providecommand{\BIBentryALTinterwordstretchfactor}{4}
\providecommand{\BIBentryALTinterwordspacing}{\spaceskip=\fontdimen2\font plus
\BIBentryALTinterwordstretchfactor\fontdimen3\font minus
  \fontdimen4\font\relax}
\providecommand{\BIBforeignlanguage}[2]{{%
\expandafter\ifx\csname l@#1\endcsname\relax
\typeout{** WARNING: IEEEtran.bst: No hyphenation pattern has been}%
\typeout{** loaded for the language `#1'. Using the pattern for}%
\typeout{** the default language instead.}%
\else
\language=\csname l@#1\endcsname
\fi
#2}}
\providecommand{\BIBdecl}{\relax}
\BIBdecl

\bibitem{Pru1964}
S.~{Pruzansky}, ``{Talker-Recognition Procedure Based on Analysis of
  Variance},'' \emph{Acoustical Society of America Journal}, vol.~36, no.~11,
  p. 2041, Jan. 1964.

\bibitem{REYNOLDS200019}
\BIBentryALTinterwordspacing
D.~A. Reynolds, T.~F. Quatieri, and R.~B. Dunn, ``Speaker verification using
  adapted gaussian mixture models,'' \emph{Digital Signal Processing}, vol.~10,
  no.~1, pp. 19--41, 2000. [Online]. Available:
  \url{https://www.sciencedirect.com/science/article/pii/S1051200499903615}
\BIBentrySTDinterwordspacing

\bibitem{ivector}
N.~Dehak, P.~J. Kenny, R.~Dehak, P.~Dumouchel, and P.~Ouellet, ``Front-end
  factor analysis for speaker verification,'' \emph{IEEE Transactions on Audio,
  Speech, and Language Processing}, vol.~19, no.~4, pp. 788--798, 2011.

\bibitem{Snyder2017}
\BIBentryALTinterwordspacing
D.~Snyder, D.~Garcia-Romero, D.~Povey, and S.~Khudanpur, ``Deep neural network
  embeddings for text-independent speaker verification,'' in \emph{Proc.
  Interspeech 2017}, 2017, pp. 999--1003. [Online]. Available:
  \url{http://dx.doi.org/10.21437/Interspeech.2017-620}
\BIBentrySTDinterwordspacing

\bibitem{xie2019}
W.~Xie, A.~Nagrani, J.~S. Chung, and A.~Zisserman, ``Utterance-level
  aggregation for speaker recognition in the wild,'' in \emph{ICASSP 2019 -
  2019 IEEE International Conference on Acoustics, Speech and Signal Processing
  (ICASSP)}, 2019, pp. 5791--5795.

\bibitem{muck2018}
H.~Muckenhirn, M.~Magimai.-Doss, and S.~Marcell, ``Towards directly modeling
  raw speech signal for speaker verification using cnns,'' in \emph{2018 IEEE
  International Conference on Acoustics, Speech and Signal Processing
  (ICASSP)}, 2018, pp. 4884--4888.

\bibitem{zhao2019robust}
F.~Zhao, H.~Li, and X.~Zhang, ``A robust text-independent speaker verification
  method based on speech separation and deep speaker,'' in \emph{ICASSP
  2019-2019 IEEE International Conference on Acoustics, Speech and Signal
  Processing (ICASSP)}.\hskip 1em plus 0.5em minus 0.4em\relax IEEE, 2019, pp.
  6101--6105.

\bibitem{shon2019voiceid}
S.~Shon, H.~Tang, and J.~Glass, ``Voiceid loss: Speech enhancement for speaker
  verification,'' \emph{arXiv preprint arXiv:1904.03601}, 2019.

\bibitem{kolboek2016speech}
M.~Kolboek, Z.-H. Tan, and J.~Jensen, ``Speech enhancement using long
  short-term memory based recurrent neural networks for noise robust speaker
  verification,'' in \emph{2016 IEEE spoken language technology workshop
  (SLT)}.\hskip 1em plus 0.5em minus 0.4em\relax IEEE, 2016, pp. 305--311.

\bibitem{chang2017robust}
J.~Chang and D.~Wang, ``Robust speaker recognition based on dnn/i-vectors and
  speech separation,'' in \emph{2017 IEEE International Conference on
  Acoustics, Speech and Signal Processing (ICASSP)}.\hskip 1em plus 0.5em minus
  0.4em\relax IEEE, 2017, pp. 5415--5419.

\bibitem{michelsanti2017conditional}
D.~Michelsanti and Z.-H. Tan, ``Conditional generative adversarial networks for
  speech enhancement and noise-robust speaker verification,'' \emph{arXiv
  preprint arXiv:1709.01703}, 2017.

\bibitem{plchot2016audio}
O.~Plchot, L.~Burget, H.~Aronowitz, and P.~Matejka, ``Audio enhancing with dnn
  autoencoder for speaker recognition,'' in \emph{2016 IEEE International
  Conference on Acoustics, Speech and Signal Processing (ICASSP)}.\hskip 1em
  plus 0.5em minus 0.4em\relax IEEE, 2016, pp. 5090--5094.

\bibitem{novotny2018use}
O.~Novotny, O.~Plchot, P.~Matejka, and O.~Glembek, ``On the use of dnn
  autoencoder for robust speaker recognition,'' \emph{arXiv preprint
  arXiv:1811.02938}, 2018.

\bibitem{meng2019}
Z.~Meng, Y.~Zhao, J.~Li, and Y.~Gong, ``Adversarial speaker verification,'' in
  \emph{ICASSP 2019 - 2019 IEEE International Conference on Acoustics, Speech
  and Signal Processing (ICASSP)}, 2019, pp. 6216--6220.

\bibitem{Peri2020}
R.~Peri, M.~Pal, A.~Jati, K.~Somandepalli, and S.~Narayanan, ``Robust speaker
  recognition using unsupervised adversarial invariance,'' in \emph{ICASSP 2020
  - 2020 IEEE International Conference on Acoustics, Speech and Signal
  Processing (ICASSP)}, 2020, pp. 6614--6618.

\bibitem{Qin2019}
\BIBentryALTinterwordspacing
X.~Qin, D.~Cai, and M.~Li, ``{Far-Field End-to-End Text-Dependent Speaker
  Verification Based on Mixed Training Data with Transfer Learning and
  Enrollment Data Augmentation},'' in \emph{Proc. Interspeech 2019}, 2019, pp.
  4045--4049. [Online]. Available:
  \url{http://dx.doi.org/10.21437/Interspeech.2019-1542}
\BIBentrySTDinterwordspacing

\bibitem{taherian2019deep}
H.~Taherian, Z.-Q. Wang, and D.~Wang, ``Deep learning based multi-channel
  speaker recognition in noisy and reverberant environments,'' in
  \emph{Interspeech}, 2019.

\bibitem{yang2019joint}
J.-Y. Yang and J.-H. Chang, ``Joint optimization of neural acoustic beamforming
  and dereverberation with x-vectors for robust speaker verification.'' in
  \emph{Interspeech}, 2019, pp. 4075--4079.

\bibitem{cai2019multi}
D.~Cai, X.~Qin, and M.~Li, ``Multi-channel training for end-to-end speaker
  recognition under reverberant and noisy environment.'' in \emph{Interspeech},
  2019, pp. 4365--4369.

\bibitem{he2021multi}
W.~He, P.~Motlicek, and J.-M. Odobez, ``Multi-task neural network for robust
  multiple speaker embedding extraction.'' in \emph{Interspeech}, 2021, pp.
  506--510.

\bibitem{wang2022spatial}
J.~Wang, Y.~Liu, B.~Wang, Y.~Zhi, S.~Li, S.~Xia, J.~Zhang, F.~Tong, L.~Li, and
  Q.~Hong, ``Spatial-aware speaker diarization for multi-channel multi-party
  meeting,'' \emph{arXiv preprint arXiv:2209.12002}, 2022.

\bibitem{kataria2020multi}
S.~Kataria, S.-X. Zhang, and D.~Yu, ``Multi-channel speaker verification for
  single and multi-talker speech,'' \emph{arXiv preprint arXiv:2010.12692},
  2020.

\bibitem{zhang2021deep}
X.-L. Zhang, ``Deep ad-hoc beamforming,'' \emph{Computer Speech \& Language},
  vol.~68, p. 101201, 2021.

\bibitem{TADRN}
\BIBentryALTinterwordspacing
A.~Pandey, B.~Xu, A.~Kumar, J.~Donley, P.~Calamia, and D.~Wang, ``{TADRN:}
  triple-attentive dual-recurrent network for ad-hoc array multichannel speech
  enhancement,'' \emph{CoRR}, vol. abs/2110.11844, 2021. [Online]. Available:
  \url{https://arxiv.org/abs/2110.11844}
\BIBentrySTDinterwordspacing

\bibitem{fasnet}
Y.~Luo, C.~Han, N.~Mesgarani, E.~Ceolini, and S.-C. Liu, ``Fasnet: Low-latency
  adaptive beamforming for multi-microphone audio processing,'' in \emph{2019
  IEEE Automatic Speech Recognition and Understanding Workshop (ASRU)}, 2019,
  pp. 260--267.

\bibitem{yang2022deep}
Z.~Yang, S.~Guan, and X.-L. Zhang, ``Deep ad-hoc beamforming based on speaker
  extraction for target-dependent speech separation,'' \emph{Speech
  Communication}, vol. 140, pp. 87--97, 2022.

\bibitem{wang2021continuous}
D.~Wang, T.~Yoshioka, Z.~Chen, X.~Wang, T.~Zhou, and Z.~Meng, ``Continuous
  speech separation with ad hoc microphone arrays,'' in \emph{2021 29th
  European Signal Processing Conference (EUSIPCO)}.\hskip 1em plus 0.5em minus
  0.4em\relax IEEE, 2021, pp. 1100--1104.

\bibitem{tac}
Y.~Luo, Z.~Chen, N.~Mesgarani, and T.~Yoshioka, ``End-to-end microphone
  permutation and number invariant multi-channel speech separation,'' in
  \emph{ICASSP 2020 - 2020 IEEE International Conference on Acoustics, Speech
  and Signal Processing (ICASSP)}, 2020, pp. 6394--6398.

\bibitem{chen2021scaling}
J.~Chen and X.-L. Zhang, ``Scaling sparsemax based channel selection for speech
  recognition with ad-hoc microphone arrays,'' \emph{arXiv preprint
  arXiv:2103.15305}, 2021.

\bibitem{liang2021attention}
C.~Liang, J.~Chen, S.~Guan, and X.-L. Zhang, ``Attention-based multi-channel
  speaker verification with ad-hoc microphone arrays,'' in \emph{2021
  Asia-Pacific Signal and Information Processing Association Annual Summit and
  Conference (APSIPA ASC)}.\hskip 1em plus 0.5em minus 0.4em\relax IEEE, 2021,
  pp. 1111--1115.

\bibitem{cai2021embedding}
D.~Cai and M.~Li, ``Embedding aggregation for far-field speaker verification
  with distributed microphone arrays,'' in \emph{2021 IEEE Spoken Language
  Technology Workshop (SLT)}.\hskip 1em plus 0.5em minus 0.4em\relax IEEE,
  2021, pp. 308--315.

\bibitem{jacob2021icassp}
P.~Tzirakis, A.~Kumar, and J.~Donley, ``Multi-channel speech enhancement using
  graph neural networks,'' in \emph{ICASSP 2021 - 2021 IEEE International
  Conference on Acoustics, Speech and Signal Processing (ICASSP)}, 2021, pp.
  3415--3419.

\bibitem{Haostgraph}
M.~Hao, J.~Yu, and L.~Zhang, ``Spatial-temporal graph convolution network for
  multichannel speech enhancement,'' in \emph{ICASSP 2022 - 2022 IEEE
  International Conference on Acoustics, Speech and Signal Processing
  (ICASSP)}, 2022, pp. 6512--6516.

\bibitem{2021arXiv211005975L}
C.~{Liang}, Y.~{Chen}, J.~{Yao}, and X.-L. {Zhang}, ``Multi-channel far-field
  speaker verification with large-scale ad-hoc microphone arrays,'' in
  \emph{Proceedings of Interspeech}, 2022, pp. 3679--3683.

\bibitem{wu2020comprehensive}
Z.~Wu, S.~Pan, F.~Chen, G.~Long, C.~Zhang, and S.~Y. Philip, ``A comprehensive
  survey on graph neural networks,'' \emph{IEEE transactions on neural networks
  and learning systems}, vol.~32, no.~1, pp. 4--24, 2020.

\bibitem{kipf2016semi}
T.~N. Kipf and M.~Welling, ``Semi-supervised classification with graph
  convolutional networks,'' \emph{arXiv preprint arXiv:1609.02907}, 2016.

\bibitem{defferrard2016convolutional}
M.~Defferrard, X.~Bresson, and P.~Vandergheynst, ``Convolutional neural
  networks on graphs with fast localized spectral filtering,'' \emph{Advances
  in neural information processing systems}, vol.~29, 2016.

\bibitem{li2015gated}
Y.~Li, D.~Tarlow, M.~Brockschmidt, and R.~Zemel, ``Gated graph sequence neural
  networks,'' \emph{arXiv preprint arXiv:1511.05493}, 2015.

\bibitem{levie2018cayleynets}
R.~Levie, F.~Monti, X.~Bresson, and M.~M. Bronstein, ``Cayleynets: Graph
  convolutional neural networks with complex rational spectral filters,''
  \emph{IEEE Transactions on Signal Processing}, vol.~67, no.~1, pp. 97--109,
  2018.

\bibitem{hamilton2017inductive}
W.~Hamilton, Z.~Ying, and J.~Leskovec, ``Inductive representation learning on
  large graphs,'' \emph{Advances in neural information processing systems},
  vol.~30, 2017.

\bibitem{velivckovic2017graph}
P.~Veli{\v{c}}kovi{\'c}, G.~Cucurull, A.~Casanova, A.~Romero, P.~Lio, and
  Y.~Bengio, ``Graph attention networks,'' \emph{arXiv preprint
  arXiv:1710.10903}, 2017.

\bibitem{zhang2018end}
M.~Zhang, Z.~Cui, M.~Neumann, and Y.~Chen, ``An end-to-end deep learning
  architecture for graph classification,'' in \emph{Proceedings of the AAAI
  conference on artificial intelligence}, vol.~32, no.~1, 2018.

\bibitem{yan2018spatial}
S.~Yan, Y.~Xiong, and D.~Lin, ``Spatial temporal graph convolutional networks
  for skeleton-based action recognition,'' in \emph{Proceedings of the AAAI
  conference on artificial intelligence}, vol.~32, no.~1, 2018.

\bibitem{shi2020point}
W.~Shi and R.~Rajkumar, ``Point-gnn: Graph neural network for 3d object
  detection in a point cloud,'' in \emph{Proceedings of the IEEE/CVF conference
  on computer vision and pattern recognition}, 2020, pp. 1711--1719.

\bibitem{shu20193d}
D.~W. Shu, S.~W. Park, and J.~Kwon, ``3d point cloud generative adversarial
  network based on tree structured graph convolutions,'' in \emph{Proceedings
  of the IEEE/CVF international conference on computer vision}, 2019, pp.
  3859--3868.

\bibitem{HuDSTGNN}
J.~Hu, X.~Lin, and C.~Wang, ``Dstgcn: Dynamic spatial-temporal graph
  convolutional network for traffic prediction,'' \emph{IEEE Sensors Journal},
  vol.~22, no.~13, pp. 13\,116--13\,124, 2022.

\bibitem{guo2019attention}
S.~Guo, Y.~Lin, N.~Feng, C.~Song, and H.~Wan, ``Attention based
  spatial-temporal graph convolutional networks for traffic flow forecasting,''
  in \emph{Proceedings of the AAAI conference on artificial intelligence},
  vol.~33, no.~01, 2019, pp. 922--929.

\bibitem{wu2019Graphwavenet}
\BIBentryALTinterwordspacing
Z.~Wu, S.~Pan, G.~Long, J.~Jiang, and C.~Zhang, ``Graph wavenet for deep
  spatial-temporal graph modeling,'' 2019. [Online]. Available:
  \url{https://arxiv.org/abs/1906.00121}
\BIBentrySTDinterwordspacing

\bibitem{wang2019stream}
X.~Wang, R.~Li, S.~H. Mallidi, T.~Hori, S.~Watanabe, and H.~Hermansky, ``Stream
  attention-based multi-array end-to-end speech recognition,'' in \emph{ICASSP
  2019-2019 IEEE International Conference on Acoustics, Speech and Signal
  Processing (ICASSP)}.\hskip 1em plus 0.5em minus 0.4em\relax IEEE, 2019, pp.
  7105--7109.

\bibitem{brody2021attentive}
S.~Brody, U.~Alon, and E.~Yahav, ``How attentive are graph attention
  networks?'' \emph{arXiv preprint arXiv:2105.14491}, 2021.

\bibitem{gao2019graph}
H.~Gao and S.~Ji, ``Graph u-nets,'' in \emph{international conference on
  machine learning}.\hskip 1em plus 0.5em minus 0.4em\relax PMLR, 2019, pp.
  2083--2092.

\bibitem{guan2021libri}
S.~Guan, S.~Liu, J.~Chen, W.~Zhu, S.~Li, X.~Tan, Z.~Yang, M.~Xu, Y.~Chen,
  C.~Liang \emph{et~al.}, ``Libri-adhoc40: A dataset collected from
  synchronized ad-hoc microphone arrays,'' in \emph{2021 Asia-Pacific Signal
  and Information Processing Association Annual Summit and Conference (APSIPA
  ASC)}.\hskip 1em plus 0.5em minus 0.4em\relax IEEE, 2021, pp. 1116--1120.

\bibitem{qin2020hi}
X.~Qin, H.~Bu, and M.~Li, ``Hi-mia: A far-field text-dependent speaker
  verification database and the baselines,'' in \emph{ICASSP 2020-2020 IEEE
  International Conference on Acoustics, Speech and Signal Processing
  (ICASSP)}.\hskip 1em plus 0.5em minus 0.4em\relax IEEE, 2020, pp. 7609--7613.

\bibitem{panayotov2015librispeech}
V.~Panayotov, G.~Chen, D.~Povey, and S.~Khudanpur, ``Librispeech: an asr corpus
  based on public domain audio books,'' in \emph{2015 IEEE international
  conference on acoustics, speech and signal processing (ICASSP)}.\hskip 1em
  plus 0.5em minus 0.4em\relax IEEE, 2015, pp. 5206--5210.

\bibitem{tan2021speech}
X.~Tan and X.-L. Zhang, ``Speech enhancement aided end-to-end multi-task
  learning for voice activity detection,'' in \emph{ICASSP 2021-2021 IEEE
  International Conference on Acoustics, Speech and Signal Processing
  (ICASSP)}.\hskip 1em plus 0.5em minus 0.4em\relax IEEE, 2021, pp. 6823--6827.

\bibitem{chung2020defence}
J.~S. Chung, J.~Huh, S.~Mun, M.~Lee, H.~S. Heo, S.~Choe, C.~Ham, S.~Jung, B.-J.
  Lee, and I.~Han, ``In defence of metric learning for speaker recognition,''
  \emph{arXiv preprint arXiv:2003.11982}, 2020.

\bibitem{du2018aishell}
J.~Du, X.~Na, X.~Liu, and H.~Bu, ``Aishell-2: Transforming mandarin asr
  research into industrial scale,'' \emph{arXiv preprint arXiv:1808.10583},
  2018.

\bibitem{anguera2007acoustic}
X.~Anguera, C.~Wooters, and J.~Hernando, ``Acoustic beamforming for speaker
  diarization of meetings,'' \emph{IEEE Transactions on Audio, Speech, and
  Language Processing}, vol.~15, no.~7, pp. 2011--2022, 2007.

\bibitem{wolf2014channel}
M.~Wolf and C.~Nadeu, ``Channel selection measures for multi-microphone speech
  recognition,'' \emph{Speech Communication}, vol.~57, pp. 170--180, 2014.

\bibitem{liang2022apsipa}
C.~Liang, J.~Chen, S.~Guan, and X.-L. Zhang, ``Attention-based multi-channel
  speaker verification with ad-hoc microphone arrays,'' in \emph{2021
  Asia-Pacific Signal and Information Processing Association Annual Summit and
  Conference (APSIPA ASC)}.\hskip 1em plus 0.5em minus 0.4em\relax IEEE, 2021,
  pp. 1111--1115.

\end{thebibliography}
	
\end{document}